\documentclass[twocolumn,preprint,nofootinbib]{aastex631}

\usepackage{amsmath}
\usepackage{graphicx}
\usepackage{color}
\usepackage{multirow}
\usepackage{etoolbox}
\usepackage{makecell}
\usepackage{comment}

\submitjournal{ApJ}

\begin{document}

\title{Hot, Retrograde Tilted MADs: Misaligned, Precessing, and Shaped by Electromagnetic Torques}

\author[0000-0001-5661-7104]{Sajal Gupta}
\affiliation{JILA and Department of Astrophysical and Planetary Sciences, University of Colorado, Boulder, CO 80309, USA}
\email{Sajal.Gupta@colorado.edu}

\author[0000-0003-3903-0373]{Jason Dexter}
\affiliation{JILA and Department of Astrophysical and Planetary Sciences, University of Colorado, Boulder, CO 80309, USA}

\begin{abstract}
Tilted accretion disks in the magnetically arrested (MAD) state may be present in X-ray binaries and active galactic nuclei such as Sgr A* and M87. We have carried out 3D global GRMHD simulations to study the evolution of these accretion flows as a function of black hole spin and misalignment angle. Prograde MADs align with the spin through a two-stage process: an initial rapid alignment phase that operates on the magnetic flux saturation timescale, followed by a slower, spin-independent phase. In contrast, retrograde MADs remain persistently misaligned regardless of the black hole spin, displaying solid-body precession at rates four times higher than weakly magnetized flows at the same spin magnitude. By deriving torque equations in ideal GRMHD and evaluating them in a frame aligned with instantaneous disk orientation, we demonstrate that electromagnetic (EM) torques always act to align the disk with the BH spin, but are countered by opposing hydrodynamic fluxes in retrograde flows. We further develop a preliminary empirical model to explain the cause of two-stage prograde alignment and discuss the possibility of alignment in the retrograde MAD. Strongly magnetized, retrograde, misaligned accretion disks provide a candidate scenario for the low-frequency quasi-periodic oscillations in black hole X-ray binaries.
\end{abstract}

\section{Introduction} \label{sec:intro}
The angular momentum of a black hole (BH) is largely shaped by its history of interactions\textemdash from mergers with other massive BHs and stars to accretion of surrounding gas \citep{Bardeen1970,King2005}. Because of these dynamic processes, the inflowing plasma and its magnetic field can often be misaligned with the BH's spin axis. In many astrophysical settings, external torques naturally induce such tilts: tidal interactions in X-ray binaries \citep[XRBs;][]{katz1980acceleration,Larwood1998,Wijers1999}, radiation pressure \citep{Pringle-radiation-II}, and stellar winds\textemdash including those from Wolf–Rayet stars near Sgr A* \citep{Schandl1994,Ghez1998,Genzel2010}\textemdash can all misalign the accretion flow relative to the BH spin axis.

While the Bardeen-Petterson effect \citep{Bardeen_Petterson} is often invoked to suggest that the inner parts of the accretion disk and BH spin should achieve mutual alignment, both numerical \citep{Nelson_Papaloizou_2000,Liska2018,Raj_2021} and analytical studies \citep{Nixon2012,Nixon2013analytical} show that this effect is primarily effective for thin, viscous disks with relatively small tilt angles. Thicker disks, however, tend to warp further away from alignment \citep{Papaloizou_Lin_1995,Ivanov_Illarionov_1996,Lubow2002}. For instance, hot, weakly magnetized accretion flows\textemdash such as those likely feeding Sgr A* \citep{Yuan_2003,EventHorizonTelescopeCollaboration_2022}\textemdash and systems undergoing super-Eddington accretion \citep{Abramowicz1980,Jiang_2019}. In such systems, alignment with the BH spin remains partial at best, with only the inner disk showing a weak tendency to align \citep{Fragile_2007,White_2019,Sajal2024}.

However, alignment dynamics differ significantly in magnetically arrested disks \citep[MADs;][]{Narayan2003,Igumenshchev2003}. Such flows are likely prevalent in systems like M87 \citep{Akiyama_2021_field_structure,Yuan_2022}, certain XRBs \citep{Bei2023}, and active galactic nuclei \citep{Mckinney_2012}. In MAD configurations, the accretion flow carries a large amount of coherent poloidal magnetic flux that accumulates and saturates near the BH \citep{Igumenshchev_2008,Begelman_2022}, leading to a formation of a relativistic jet \citep{Blandford_Znajek_1977,Blandford_Payne_1982,Tchekhovskoy2011,Narayan2012}. 

\cite{Mckinney_tilted_2013} suggested that in tilted MADs, this relativistic jet exerts an asymmetric force on the disk \citep{Polko2016}, promoting alignment with the BH spin even for fast-spinning BHs undergoing extreme misalignment ($\sim 90^\circ$). They coined the term `magneto-spin alignment' to describe this mechanism, and suggested that it may operate for moderately rotating BHs, with spin as low as $a \gtrsim 0.2$ for thick MADs and $a \gtrsim 0.5$ for thinner ones. However, questions remain about how the jet can exert the torque needed for alignment, as modeled by \cite{Polko2016}, if it is emitted perpendicular to the average disk plane. As for the jet to drive alignment, it must push on the disk asymmetrically, implying that it cannot be strictly perpendicular. Yet, recent simulations by \cite{Liska2018jetprecess,Chatterjee2020} indicate that jets from strongly magnetized flows may indeed align closely with the disk's angular momentum vector, potentially challenging this proposed alignment mechanism.

These contrasting findings raise fundamental questions about the physical processes underlying disk alignment in MADs. Additionally, \cite{Chatterjee2023misaligned} studied hot, tilted MADs and found that retrograde disks, in contrast to \cite{Mckinney_tilted_2013}'s results, do not align with the BH axis of symmetry. This discrepancy suggests two pressing questions: why do retrograde disks remain misaligned in these simulations, and what factors in \cite{Mckinney_tilted_2013} study led to alignment across a wide range of tilt angles? 

\begin{figure*}[t!]
\centering
\includegraphics[width=\textwidth]{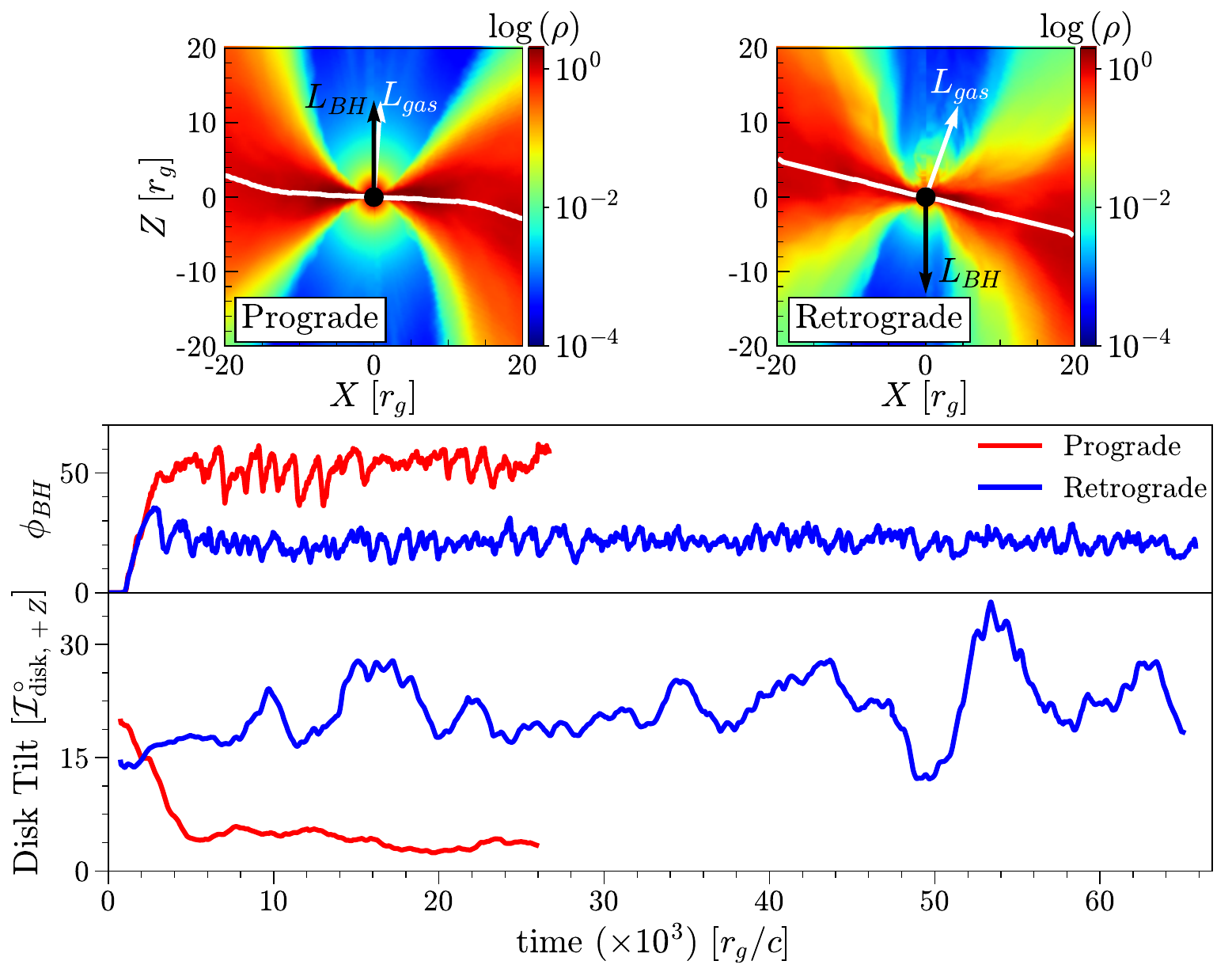}
\caption{\small{\textbf{Top panel:} Poloidal slice snapshots of density ($\rho$) for high-spin prograde ($a \approxeq 0.94$) and retrograde ($a \approxeq -0.94$) cases, captured during their respective flux saturation time. Both flows are initially tilted by $16^\circ$ from the $+ z$ axis (see equation \ref{eq:diskinclinationeq}) The white line denotes the averaged disk plane, and black and white arrows indicate the directions of the BH and disk angular momenta, respectively. \textbf{Middle panel:} Evolution of normalized magnetic flux threading the event horizon. \textbf{Bottom panel:} Disk tilt evolution from the event horizon to $20\,r_g$. The figure demonstrates that prograde flows align with the BH spin axis, whereas retrograde flows remain misaligned. Here, for comparison to the prograde case, the tilt evolution for retrograde flow was shown relative to $+z$-axis.}}
\label{fig:density_plot_and_flux_and_disk_inclination_comparison}
\end{figure*}

In this paper, we argue that jets do, in fact, emit nearly along the disk’s rotational axis, and it is primarily the magnetic stresses\textemdash particularly the radial-polar magnetic tension\textemdash that drive alignment in prograde flows. The structure of this paper is as follows: In Section \ref{sec:Methods}, we outline the governing equations and simulation setup for our ideal, general relativistic magnetohydrodynamic (GRMHD) simulations. In section \ref{sec:section of Alignment and Misalignment of Prograde and Retrograde MADs}, we analyze the tilt profiles of both the jet and the disk, showing that while retrograde flows remain misaligned, prograde flows align in two distinct phases: a rapid phase that concludes once the magnetic flux threading the BH saturates, followed by a slower alignment phase, as illustrated in Figure \ref{fig:density_plot_and_flux_and_disk_inclination_comparison}. Section~\ref{sec:Matter, Electromagnetic and Gravitational Torque} examines the relative contributions of physical process in disk alignment, showing that magnetic torques generally act to reduce the disk–spin misalignment in both prograde and retrograde flows. An empirical scaling for the alignment torque and a two-stage picture of prograde evolution is described in section \ref{subsec:Empirical_model_two_stages_alignment}. Section \ref{sec:Solid Body Precession of the Retrograde MADs} showcases the near rigid-body precession of retrograde disks, and in section \ref{sec:discussion}, we address the alignment potential for retrograde flows and discuss the retrograde precession of prograde flows. Finally, Section \ref{sec:conclusion} concludes our findings.

\section{Methods} \label{sec:Methods}
\subsection{Governing Equations} \label{sec:Governing equations}
We use the publicly available HARMPI code \citep{tchekhovskoy2019harmpi} (a parallel, 3D version of HARM \citep{Gammie_2003, Noble_2006}) to solve the idealized GRMHD equations of motion written in the following form:
\begin{align}\label{eq:GRMHD1}
    \partial_t \left(\sqrt{-g}\rho u^t\right) + \partial_j \left(\sqrt{-g}\rho u^j\right) &= 0, \\ \label{eq:GRMHD2}
    \partial_t \left(\sqrt{-g}T^t_\mu\right) + \partial_j \left(\sqrt{-g}T^j_\mu\right) - \sqrt{-g}T^\kappa_\lambda \Gamma^\lambda_{\mu \kappa} &= 0, \\ \label{eq:GRMHD3}
    \nabla_\nu \left(\mbox{*} F^{\mu \nu}\right) &= 0,
\end{align}
where $\rho$ is the fluid frame density, $u^\mu$ is the contravariant 4-velocity, $g \equiv \rm det(g_{\mu\nu})$ is the metric determinant, and $\Gamma^\lambda_{\mu \kappa}$ are the connection coefficients. The Greek alphabets are used to represent four-vector, whereas Latin alphabets describe spatial components. We describe the space-time of rotating BH using spherical polar Kerr-Schild coordinates \citep{Font_1998} with a metric signature ($- $,+,+,+). Here, $T^\mu_\nu$ and $\mbox{*} F^{\mu \nu}$ are the ideal MHD stress-energy tensor and dual electromagnetic tensor, respectively, defined as
\begin{align}\label{eq:stress-energy tensor}
    T^{\mu}_{\nu} &= \left(\rho h + b^2\right)u^\mu u_\nu + \left(p_{gas} + p_{mag}\right)\delta^{\mu}_{\nu} - b^\mu b_\nu, \\
    \mbox{*} F^{\mu \nu} &= b^\mu u^\nu - b^\nu u^\mu, 
\end{align}
where $p_{gas}$ is the gas pressure, $h = 1 + \gamma p_{gas}/(\rho (\gamma -1))$ is the specific enthalpy, $\gamma = 5/3$ is the adiabatic index, $b^\mu$ is the contravariant fluid-frame magnetic 4-field, and $p_{mag} \equiv b^{\mu}b_{\mu}/2 = b^2/2 $ is the magnetic pressure.

\begin{deluxetable*}{|c|c|c|c|c|c|c|c|c|c|}[t!]
\tablecolumns{10}
\tablewidth{1.0\textwidth} 
\tablecaption{List of the simulations presented in this work.} 
\tablehead{\colhead{BH spin} & \colhead{\makecell{$\mathcal{I}_{\text{disk},\text{BH}}^{\circ}$ \\ (t=0)}} & \colhead{\makecell{$r_{max}$ \\ $[r_g]$}} & \colhead{\makecell{Resolution \\ $N_r \times N_\theta \times N_\phi$}} & \colhead{\makecell{$t_{\text{sim}}$ \\ \textbf{[$10^3\,r_g/c$]}}} & \colhead{\makecell{\textbf{$\langle \phi_{BH} \rangle$}}} & \colhead{\makecell{$t_{\text{rapid}}$ \\ \textbf{[$10^3\,r_g/c$]}}}
& \colhead{\makecell{$\langle \mathcal{I}^{\circ}_{\text{disk},\text{BH}} \rangle$ \\ ($10\,r_g$)}} & \colhead{\makecell{$\langle \mathcal{I}^{\circ}_{\text{jet}} \rangle$ \\ ($10\,r_g$)}} & \colhead{Summary} }
\startdata
\hline
\multicolumn{10}{|c|}{Prograde MADs} \\
\hline
    0.3 & 16$^\circ$ & 25 & $320 \times 256 \times 160$ & 52.7 & 59.3 & 0.65 & 9.1$^\circ$ & 9.3$^\circ$ & Aligning \\
    0.5 & 16$^\circ$ & 25 & $320 \times 256 \times 160$ & 51.7 & 59.6 & 0.625 & 7.4$^\circ$ & 7.7$^\circ$ & Aligning \\
    0.5 & 30$^\circ$ & 25 & $320 \times 256 \times 160$ & 29.2 & 61.7 & 0.625 & 12$^\circ$ & 10.4$^\circ$ & Aligning \\
    0.75 & 16$^\circ$ & 25 & $320 \times 256 \times 160$ & 25.2 & 59.3 & 0.59 & 6$^\circ$ & 6.3$^\circ$ & Aligning \\
    0.9375 & 16$^\circ$ & 25 & $320 \times 256 \times 160$ & 26.7 & 52.7 & 0.54 & 5$^\circ$ & 6.1$^\circ$ & Aligned \\
\hline
\multicolumn{10}{|c|}{Retrograde MADs} \\
\hline
   -0.9375 & 164$^\circ$ & 27.125 & $320 \times 256 \times 160$ & 65.7 & 21.1 & – & 160.3$^\circ$ & 163.2$^\circ$ & \makecell{Persistently \\ misaligned} \\
\hline
\multicolumn{10}{|c|}{Retrograde MADs (Lower resolution test runs)} \\
\hline
   -0.1 & 150$^\circ$ & 25.2 & $160 \times 128 \times 80$ & 71.4 & 51.7 & – & 147.5$^\circ$ & 150.8$^\circ$ & \makecell{Persistently \\ misaligned} \\
   -0.5 & 150$^\circ$ & 26.1 & $160 \times 128 \times 80$ & 34.1 & 33.9 & – & 148.2$^\circ$ & 155.3$^\circ$ & \makecell{Persistently \\ misaligned} \\
   -0.9375 & 150$^\circ$ & 27.125 & $160 \times 128 \times 80$ & 80.5 & 20.4 & – & 150.1$^\circ$ & 166.4$^\circ$ & \makecell{Persistently \\ misaligned} \\
\enddata
\label{tab:MAD_models}
\tablecomments{\small{The first column indicates the BH dimensionless spin, where positive/negative values correspond to BH orientation along the $\pm z$-axis. The second column shows the initial tilt of the disk relative to the spin axis. The third, fourth, and fifth columns provide the location of the pressure maximum, resolution, and final simulation time, respectively. The sixth column presents the time-averaged MAD parameter. The seventh column marks the end of the rapid alignment phase in prograde flows, coinciding with the flux saturation timescale. The eighth and ninth columns report the evolved relative tilt between the BH spin axis and the disk and jet at $r = 10\,r_g$, respectively. The final column summarizes the disk alignment status at $r = 10\,r_g$: ‘aligning’ indicates ongoing alignment at the end of the simulation, while ‘aligned’ or ‘persistently misaligned’ indicates that the disk has reached a stable configuration.}}
\vspace*{-0.75cm}
\end{deluxetable*}

\vspace*{-0.5cm}
\subsection{Simulation Setup}
We simulate tilted MADs for seven BH spin values ($a$), ranging from $a=-0.9375$ to $a=$ $0.9375$. The models were run at two resolutions (five high-resolution and two low-resolution); see Table \ref{tab:MAD_models} for details. The dimensionless spin parameter $a$ is defined as $-1 < a \equiv~ c\mathbf{L}_{BH}/GM_{BH}^2 < 1$, and $\mathbf{L}_{BH}$ represents the angular momentum vector of a BH. Each simulation initializes an isentropic hydro-equilibrium torus \citep{Fishbone1976} with inner radius at $r_{in}=12\,r_g$, where $r_g \equiv GM_{BH}/c^2$ denotes one gravitational radii. For prograde spins ($a > 0$), the pressure maximum is fixed at $r_{max} = 25\,r_g$. Whereas, for retrograde spins ($a < 0$), $r_{max}$ is adjusted within $25 - 27\,r_g$. This adjustment is made to keep the radial extent of the tori (specifically, the outer radius $r_{out}$) approximately similar across same magnitudes of spin. Table \ref{tab:MAD_models} lists the inner radius and pressure maximum for each simulation.

The torus is then rotated about the BH's equatorial plane such that its initial angular momentum lies in the $x$-$z$ plane. We initialize tilt angles of $16^\circ$ and/or $30^\circ$ in our simulations, allowing us to investigate the effect of the initial misalignment without the disk entering the polar region. A single poloidal magnetic field loop threads the torus, initialized to saturate magnetic flux on the BH, with the maximum plasma-beta parameter ($\beta = p_{gas}/p_{mag}$) of $100$.To verify that key diagnostics (e.g. average disk inclination and magnetic-flux evolution) do not depend on whether one rotates the disk or flips the BH spin axis, we carried out symmetry tests for a $16^\circ$ misaligned disk around high-spin BHs ($a=\pm0.9375$). These tests confirm identical results under the two equivalent initialization schemes and are described in Appendix \ref{sec:Appendix_Symmetry and Tilt Consistency in Prograde and Retrograde MADs}.

We evolve all simulations in modified spherical-polar Kerr-Schild coordinates ($r,\theta,\phi$). Our primary set\textemdash comprising both prograde and high spin retrograde \textemdash uses a resolution of $320\times256\times160$ cells (see Table \ref{tab:MAD_models}). To explore how spin and tilt trends might extend beyond this core grid, we additionally perform three lower-resolution retrograde test runs ($160\times128\times80$ cells) at $a=-0.1$, $-0.5$, and $-0.9375$ (the latter with a $30^\circ$ tilt). The computational domain spans from $0.88 r_H$ to $10^5\,r_g$ in the radial direction, from $0$ to $\pi$ in the polar direction, and from $0$ to $2\pi$ in the azimuthal direction, where $r_H = r_g(1 + \sqrt{1 - a^2})$ represents BH event horizon. The outer radial boundary is extended to $10^5\,r_g$ using a superexponential radial grid \citep{Ressler2017}. We employ outflowing boundary conditions at the inner and outer radial boundaries, periodic in azimuth direction, and reflecting at the polar axes. While a reflective polar boundary can introduce additional dissipation near the axis \citep{Liska2018jetprecess}, the overall dynamics and alignment of the prograde disks in our simulations are consistent with previous studies that used transmissive BCs \citep{Mckinney_tilted_2013,Chatterjee2023misaligned}.

\begin{figure*}[t!]
\centering
\includegraphics[width=\textwidth]{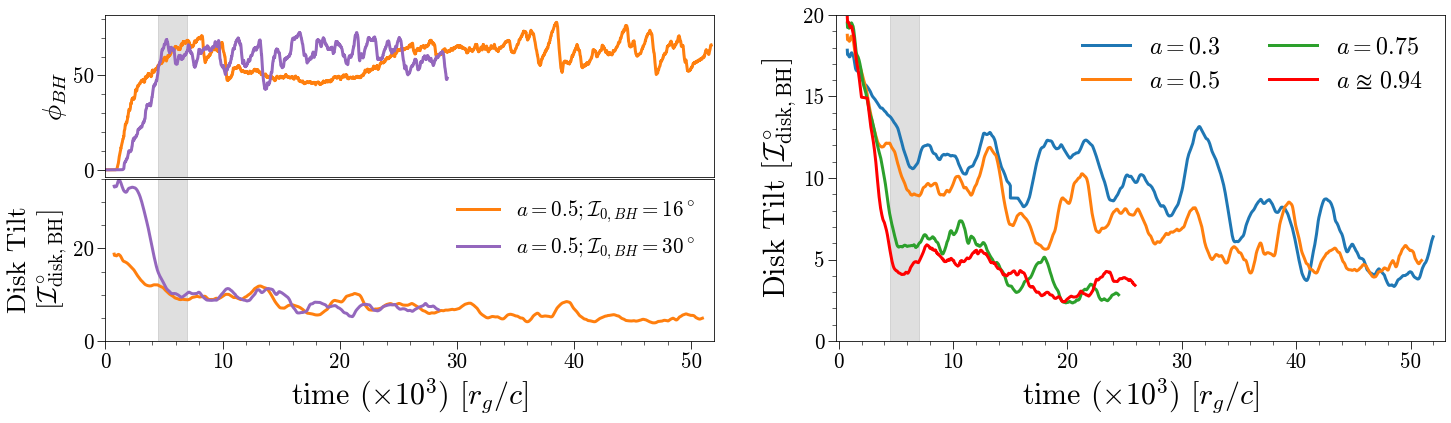}
\caption{\small{\textbf{Left column:} Top and bottom panels show the MAD parameter and the moving averages of disk tilt evolution over a timescale of $1500\,r_g/c$, from the event horizon to $20\,r_g$, for simulated $a = 0.5$ prograde cases (colors). \textbf{Right column:} Temporal evolution of disk tilt at different BH spins for prograde flows, each initially tilted with $16^\circ$ (color-coded). The figure demonstrates that disk alignment with the BH spin occurs in two distinct stages: a rapid phase followed by a slower alignment phase. The transition, highlighted in gray, aligns with the flux saturation timescale (see Table \ref{tab:MAD_models}).}}
\label{fig:dimensionless_flux_and_disk_inclination_Prograde_MAD}
\end{figure*}
We evolve the simulations until the inner disk ($r \leq 20\,r_g$) reaches a quasi-steady state with no significant changes in its average tilt. Table \ref{tab:MAD_models} lists the final simulation time for each run. Following \cite{Penna2010}, we define the shell-averaged quantities with density weight as:
\begin{equation}
    \langle x \rangle_{\rho}(r,t) = \frac{\int_{\theta}\int_{\phi} x~\rho~ dS_r}{\int_{\theta}\int_{\phi} \rho~ dS_r},
\end{equation}
where $dS_r = \sqrt{-g} d\theta d\phi$ is the area element of the r = constant surface. In addition, for our torque analysis, we define the area integral as:
\begin{equation}
\label{eq:surfaceintegral}
   [x]_{S_r}(r,t) = \int_{\theta}\int_{\phi} x dS_r
\end{equation}

\section{Alignment and Misalignment of Prograde and Retrograde MADs}
\label{sec:section of Alignment and Misalignment of Prograde and Retrograde MADs}

\subsection{Accretion flow}
The magnetic flux saturation at the horizon is often expressed in terms of a dimensionless quantity, $\phi_{BH}$, known as the ``MAD parameter''. This parameter is defined as the ratio of the spherical magnetic flux ($\Phi_{\text{sph},\text{BH}}$) threading the BH to the square root of the mass accretion rate ($\dot{M}$) at the horizon. The top panel in left column of Figure \ref{fig:dimensionless_flux_and_disk_inclination_Prograde_MAD} shows the time evolution of the MAD parameter for $a = 0.5$ runs, where we define the spherical magnetic flux and rest-mass accretion rate as:
\begin{align}
\Phi_{\text{sph}} &= \frac{1}{2} \int_{\theta=0}^{\theta = \pi} \int_{\phi=0}^{\phi=2\pi} \sqrt{4 \pi} |B^r| dS_r, \\
\dot{M} &= - \int_{\theta=0}^{\theta = \pi} \int_{\phi=0}^{\phi=2\pi} \rho u^r dS_r
\end{align}
We find that $\phi_{\text{BH}} \equiv \Phi_{H}/\sqrt{\langle\dot{M}_{\text{BH}}\rangle_{\tau} r_g^2 c}~;\Phi_H = \Phi_{\text{sph}}\big|_{r=r_H}$ saturates to a constant value across all the prograde simulations (see Figure \ref{fig:density_plot_and_flux_and_disk_inclination_comparison} for the high spin case, and Table \ref{tab:MAD_models} for saturated values), indicating that the magnetic flux reaches a steady equilibrium state. Here, $\langle\dot{M}_{BH}\rangle_{\tau}$ denotes the accretion rate at horizon, which is smoothed using a Gaussian function with a width of $\tau = 1500\,r_g/c$ to highlight long-term trends.

The bottom panel in left column of Figure \ref{fig:dimensionless_flux_and_disk_inclination_Prograde_MAD} display the time evolution of the tilt of the inner disk, measured from the horizon to $20\,r_g$, for $a = 0.5$ prograde runs. We compute the disk tilt from the BH spin axis as: 
\begin{equation}
\label{eq:diskinclinationeq}
    \mathcal{I}_{\text{disk},\text{BH}}(r,t) \equiv \cos^{-1}\left(\frac{\mathbf{L}_{BH}\cdot\langle\mathbf{L_{\text{gas}}}\rangle_{\rho}}{|\mathbf{L}_{BH}||\langle\mathbf{L_{\text{gas}}}\rangle_{\rho}|}\right)
\end{equation}
where $\mathbf{L_{\text{gas}}}$ represents the disk’s angular momentum and $\mathbf{L}_{BH} = a \hat{z}$ denotes the BH spin in natural units. The calculation of $\mathbf{L_{\text{gas}}}$\footnote{We adopt extrinsic spin formulation (gauge-dependent with reference point at the BH center; \cite{Mewes2016NumericalRS}), rather than the intrinsic formulation \citep{Fragile_2007,Mckinney_tilted_2013}; while all definitions yield consistent tilt/twist in the disk body, near the horizon strong curvature render the angular momentum decomposition coordinate-dependent.} follows the method outlined in \cite{Sajal2024}:
\begin{equation}
\label{eq:diskangmomcompeq}
    (\mathbf{L_{\text{gas}}})_{\hat{k}} = \epsilon_{ijk} r^{\hat{j}}~(T^t~_{\hat{k}})_{\text{MA}}
\end{equation}
where the subscript ``MA'' signifies the matter contribution to the stress-energy tensor \citep{Gammie_2003}, $\epsilon_{ijk}$ is the three-dimensional Levi-Civita symbol, and hats indicate Cartesian coordinates mapped to spherical coordinates in the conventional manner.

The bottom left panel of Figure \ref{fig:dimensionless_flux_and_disk_inclination_Prograde_MAD} reveal several key insights. First, the tilt of the inner disk decreases steadily over time, approaching zero with the spin axis. This alignment trend persists across all prograde simulations, regardless of initial tilt angle or spin, as shown in the right panel of Figure \ref{fig:dimensionless_flux_and_disk_inclination_Prograde_MAD}. Next, comparing the solid purple and red curves for same spin ($a = 0.5$), we observe that the initial alignment rate is higher for the $30^\circ$ case than for the $16^\circ$ tilt, despite both cases reaching similar MAD parameter. This implies that the alignment torque initially scales with the tilt angle. However, once a residual tilt of $\approx 8^\circ$ is reached, both systems align at nearly identical rates, suggesting a common late-stage alignment mechanism independent of initial tilt.

The temporal evolution of the disk tilt reveals two distinct alignment phases: a rapid alignment phase preceding magnetic flux saturation and a subsequent slow alignment phase, as evident in the right panel of Figure \ref{fig:dimensionless_flux_and_disk_inclination_Prograde_MAD}. The transition between these phases coincides with flux saturation, as inferred by comparing the tilt evolution (bottom) to flux growth (top). During the rapid phase, stronger spins drive a steeper decline in tilt, highlighting a strong spin dependence. This dependence diminishes in the slow phase, where alignment rate decreases significantly and converge across spins (e.g., $a = 0.3$ vs $a = 0.5$). We provide a plausible explanation for this dichotomy in Section \ref{subsec:Empirical_model_two_stages_alignment}. The highest spin case ($a \approxeq 0.94$) achieves nearly complete alignment, with the residual tilt of $\approx 3^\circ$ likely due to numerical errors \citep{Sajal2024}.

\begin{figure*}[t!]
\centering
\includegraphics[width=\textwidth]{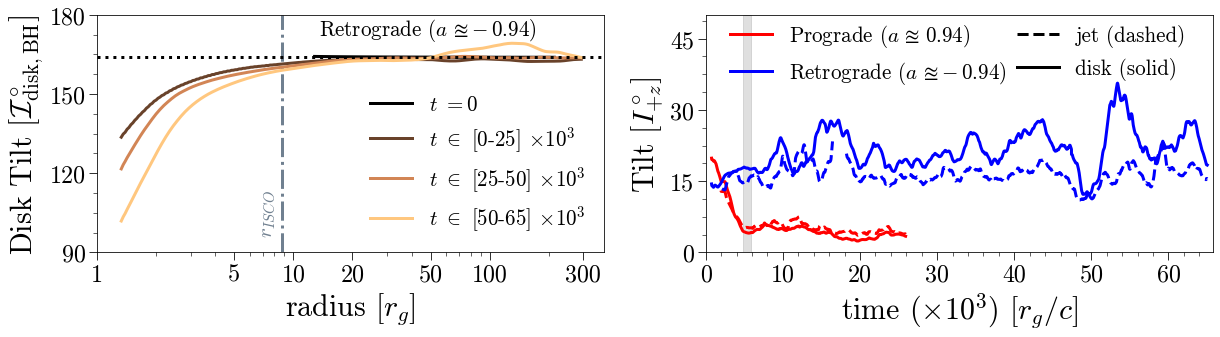}
\caption{\small{\textbf{Left column:} Density-weighted, shell-averaged radial profiles of disk tilt for the high-spin retrograde flow ($a \approxeq -0.94, \mathcal{I}^\circ_{0,BH} = 164^\circ$), averaged over various time intervals (color-coded, in units of $10^3\,r_g/c$). Within the ISCO the gas progressively shifts toward alignment with the spin axis, but never achieves complete alignment. \textbf{Right Column:} Comparing time evolution of the jet tilt (dashed) and disk tilt (solid) within the $r \in [r_H, 20\,r_g]$ region for high-spin prograde (red) and retrograde (blue) flows. The prograde jets emit perpendicular to the disk midplane, aligning at a similar rate as the disk. In contrast, high-spin retrograde jet are positioned between the disk and the BH rotation axis.}}
\label{fig:radial_profile_of_disk_inclination_Retrograde_MADs_and_jet_inclination}
\end{figure*}

In contrast, retrograde flows never fully align, even after magnetic flux saturation. This is illustrated in Figure \ref{fig:density_plot_and_flux_and_disk_inclination_comparison} for high-spin retrograde where we show the time evolution of the MAD parameter (middle panel) and disk tilt (bottom panel) relative to $+z$ axis. The disk tilt relative to the spin axis for retrograde runs is given by $\mathcal{I}^\circ_{disk,BH} = 180^\circ - \mathcal{I}^\circ_{disk,+z}$. Although high-spin retrograde does not align, its tilt relative to spin axis sometimes briefly drops by 15-20\%, lasting only for $\sim 2500\,r_g/c$ before swinging back toward the initial misalignment of $164^\circ$. The radial tilt profile in the left column of Figure \ref{fig:radial_profile_of_disk_inclination_Retrograde_MADs_and_jet_inclination} shows that inside the ISCO the disk persistently approaches orthogonality, while at larger radii it maintains its original misalignment. We observed that all of our simulated retrograde flows remained misaligned (see Table \ref{tab:MAD_models}). Even the low-resolution $a=-0.1$ case\textemdash despite sustaining $\sim2.5\times$ more magnetic flux\textemdash fails to achieve lasting alignment. This indicates that alignment depends not solely on MAD parameter but also on an intricate interplay between BH spin and magnetic field strength.

\begin{figure*}[t!]
\centering
\includegraphics[width=\textwidth]{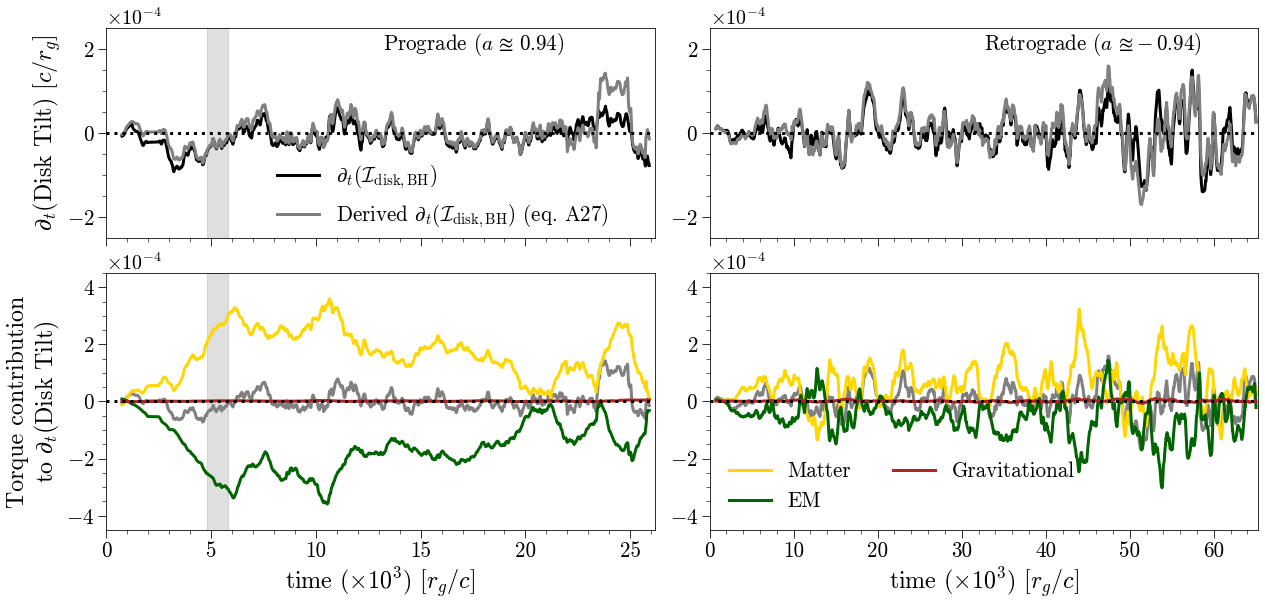}
\caption{\small{Torque analysis in disk-aligned frame for high-spin prograde (left column) and retrograde (right column) flows, each initially tilted by $16^\circ$ relative to the $+ z$-axis. \textbf{Top:} Comparing rate of change of disk tilt (black) within $r \in [15,25]\,r_g$ against our derived angular momentum evolution equations (gray; Appendix \ref{sec:Appendix_Torque equations_in_DF}) in disk-aligned frame. \textbf{Bottom:} torque contributions \textemdash gravitational (red) is negligible, EM (green) is negative for both prograde and retrograde flows, driving alignment with the spin axis. The matter torque (yellow) is always found opposing the magnetic forces, sustaining misalignment in retrograde flows.}}
\label{fig:torque_contribution_to_rate_of_change_of_tilt}
\end{figure*}

\subsection{Direction of the Relativistic Jet}
We measured the tilt of the relativistic jet using the method proposed by \cite{Liska2018jetprecess} (with a modification detailed in Appendix \ref{sec:appendix_jet_inclination}) and compare them to disk tilts in the right panel of Figure \ref{fig:radial_profile_of_disk_inclination_Retrograde_MADs_and_jet_inclination} for high-spin prograde and retrograde case. Both quantities are tracked from the horizon to $20\,r_g$. In prograde flows, the jet and disk align synchronously with minimal lag, indicating jet emission along the disk’s angular momentum axis. In contrast, relativistic jets in the retrograde flow remain misaligned relative to the spin axis, mirroring the disk’s orientation.

\section{Torque Analysis}
\label{sec:Matter, Electromagnetic and Gravitational Torque}
\subsection{Direction of the Electromagnetic Torque: What is it pointing towards?}

To investigate the dynamical evolution of tilted MADs, it is crucial to study the various processes that can torque the surrounding gas. To this end, we have derived a set of equations theta govern the total angular momentum evolution in the ideal GRMHD framework\textemdash excluding contributions from any external forces or viscosity. By ``total'', we refer to the combined angular momentum of both the gas and the electromagnetic (EM) fields. The detailed derivation of these equations is presented in Appendix \ref{sec:Appendix_Torque equations} and \ref{sec:Appendix_Torque equations_in_DF}, where we outline the three primary sources of angular momentum flux: 
\begin{enumerate} 
\item Hydrodynamic (matter/inertial) torque, which governs angular momentum exchange between neighboring rings via the radial flux of gas 
\item EM torque, driven by large-scale magnetic stresses 
\item Gravitational torque due to curved spacetime. 
\end{enumerate}

In this analysis, we reasonably assume that the angular momentum contribution from the magnetic fields is negligible compared to the gas. Consequently, rather than focusing solely on the angular momentum derived from the matter component of the stress-energy tensor, we analyze the evolution of the total angular momentum vector to study disk alignment. Appendix \ref{sec:Appendix_Tilt using total angular momentum} provides detailed justification for this assumption. Additionally, to further explore the magnetic field's impact on average disk inclination, we perform radial surface integration (equation \ref{eq:surfaceintegral}) without applying any density weighting.

Figure \ref{fig:torque_contribution_to_rate_of_change_of_tilt} presents the moving averages of the rate of change of disk tilt, measured from $15\,r_g$ to $25\,r_g$ in disk-aligned frame (see \ref{sec:Appendix_Torque equations_in_DF}), along with the contributions of different torques over a time scale of $1500\,r_g/c$. The high-spin prograde ($a \approxeq 0.94$, tilt $= 16^\circ$, left panel) and retrograde ($a \approxeq -0.94$, tilt $= 164^\circ$, right panel) cases are shown for comparison. We selected $r \in [15,25]\,r_g$ radial shell, rather than a region closer to the horizon, to avoid large numerical fluctuations and to be coherent with the rest of paper. However, we do find the following observations to be consistent for $r \geq 5\,r_g$.

The top panels of Figure \ref{fig:torque_contribution_to_rate_of_change_of_tilt} assess the accuracy of the angular momentum equations in post-processing. There is a small discrepancy between the calculated rate from the radial tilt profile and that predicted by the torque equations. It is because we did not compute the rates on the fly. The bottom panels of Figure \ref{fig:torque_contribution_to_rate_of_change_of_tilt} display the individual contributions of different torques\textemdash matter, EM, and gravitational. Several key observations arise from these plots. First, we observe that the contribution from gravitational torque is negligible relative to both the matter and EM torques, as expected from Lense-Thirring (LT; \cite{Lense-Thirring}) torque being perpendicular to disk angular momentum vector. 

More critically, we find that the EM torque contribution is negative in both the prograde and retrograde cases, and it is approximately twice as strong in the former as in the latter. This indicates that the EM fields torque the disk to align with the spin axis. In the retrograde case, this is particularly significant because the disk's rotational axis could, in principle, become either parallel or anti-parallel. The matter torque, set by the disk’s inertia, is observed to always oppose the EM contribution and appears to adjust its magnitude in response to EM torque. In prograde disk, this opposition slows but does not prevent alignment. In retrograde flows, however, the inertial response nearly cancels the EM torque, maintaining a quasi-steady misalignment despite the negative EM signature.

\begin{figure*}[t!]
\centering
\includegraphics[width=\textwidth]{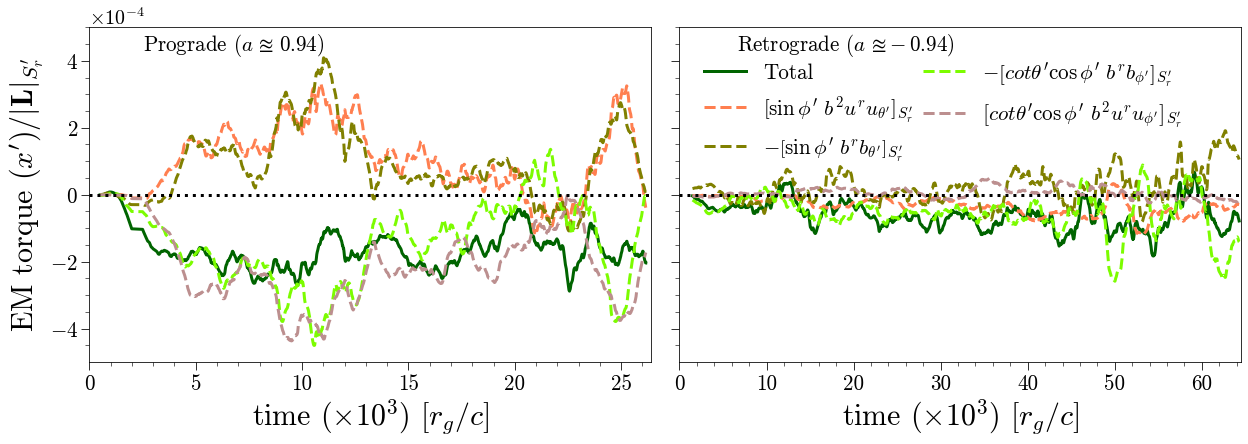}
\caption{\small{Electromagnetic contribution to the $x^\prime$–component of $\partial_r G_{x^\prime}^{\rm EM}/\big|[\boldsymbol L]\big|_{S_r}$, for high–spin prograde (left) and retrograde (right) MADs. All terms are evaluated in the disk-aligned frame for region $15\le r/r_g\le 25$ with polar caps within $10^\circ$ of disk polar axes are excluded. The solid dark green curve shows the net EM contribution to alignment torque. The dashed curves represent decomposition of net EM alignment torque into individual components as expressed in equation \ref{eq:GxprimetorqueEMpaper}. Prograde runs show a net negative torque dominated by radial-azimuthal magnetic channels, whereas all magnetic channels are substantially weaker in retrograde case.}}
\label{fig:xprime_component_of_EM_torque}
\end{figure*}

\subsection{What Physical Processes Produce EM Torque and drive alignment?}
To identify which electromagnetic stresses drive alignment, we decompose the $x^\prime$-projected EM torque per shell. A caveat is that the tilt rate also contains a geometric shear correction (a Coriolis-like force, see \ref{sec:Appendix_Torque equations_in_DF}) arising purely from the spatial variation of the disk-aligned basis across radius, i.e. from the geometry of the annulus definition rather than a distinct channel of angular momentum flux. In practice, the shear term becomes numerically large only near the disk's polar axes ($\theta^\prime \to 0,\pi$). We therefore exclude the polar caps $\theta^\prime \leq 10^\circ$ and $\theta^\prime \geq 170^\circ$, under which we find $\mathcal{S}_{x^\prime} \ll \partial_r(G_{x^\prime})$ for $r \geq 10$ and hence $\partial_t(\mathcal{I}_{+z}) \approx \frac{\partial_r G_{x^\prime}^{\text{total}}}{\left|\left[\mathbf{L}\right]_{S_r}\right|}$.

Breaking up the EM contribution of this alignment torque at a fixed radius yields: 
\begin{equation}
\label{eq:GxprimetorqueEMpaper}
    \begin{split}
    &\partial_r(G_{x^\prime}^{\text{EM}}) \equiv \partial_r \left[\sin{\phi^\prime} ~T^r_{\theta^\prime,\text{EM}} + cot{\theta^\prime} \cos{\phi^\prime} ~T^r_{\phi^\prime,\text{EM}} \right]_{S^\prime_r} \\ 
     &= \partial_r \left[ \sin{\phi^\prime}~ b^2 u^{r} u_{\theta^\prime} \right]_{S^\prime_r} - \partial_r \left[ \sin{\phi^\prime}~ b^{r} b_{\theta^\prime} \right]_{S^\prime_r} ~+ \\
     &\partial_r \left[ cot{\theta^\prime} \cos{\phi^\prime}~ b^2 u^{r} u_{\phi^\prime} \right]_{S^\prime_r} - \partial_r \left[ cot{\theta^\prime} \cos{\phi^\prime}~ b^{r} b_{\phi^\prime} \right]_{S^\prime_r}
    \end{split}
    \raisetag{32.5pt}
\end{equation}
where we used the fact that $r^\prime = r$. We will refer to the first and third terms as advective EM-inertia channels and to the second and fourth as Maxwell-tension channels relative to the disk orientation.

The left and right panels of Figure \ref{fig:xprime_component_of_EM_torque} show these contributions, normalized by $|\left[\boldsymbol{L}\right]_{S^\prime_r}|$, for high-spin prograde (left) and retrograde (right) MADs in $15\,r_g \leq r \leq 25\,r_g$ (polar caps $\theta^\prime \leq 10^\circ$ and $\theta^\prime \geq 170^\circ$ excluded). The thick green curve, (sum of all EM terms; equation \ref{eq:GxprimetorqueEMpaper}) is negative in both cases, i.e. the net magnetic forces contributes to alignment. In prograde systems, alignment is driven by large-scale radial-azimuthal magnetic tension, along with comparable $r\phi^\prime$ EM-inertia (advective) term. By contrast, the radial-polar magnetic forces \emph{relative to the disk orientation} enter with the opposite sign and partially offset alignment. In retrograde flows all components are weaker, intermittent alignment episodes are driven mainly by the radial-azimuthal correlation term, while the other channels fluctuate about zero with small amplitude. This contrast highlights that alignment dynamics depend sensitively on the relative orientation of the flow, while the detailed mechanisms setting the signs of individual forces remain uncertain.

\subsection{An empirical model for the alignment torque: evidence for two-stage prograde disk evolution}
\label{subsec:Empirical_model_two_stages_alignment}

Prograde disks and jets align in two phases (Figures \ref{fig:dimensionless_flux_and_disk_inclination_Prograde_MAD}, \ref{fig:radial_profile_of_disk_inclination_Retrograde_MADs_and_jet_inclination}): rapid alignment operating on the flux saturation timescale, followed by a gradual residual tilt reduction. One might expect the alignment rate to peak when the flux saturates, as this marks the disk's transition to a magnetically arrested state, but we instead observe a decline. This counterintuitive behavior may be due to weakening EM torque as the disk partially aligns. 

\begin{figure*}[t!]
\centering
\includegraphics[width=\textwidth]{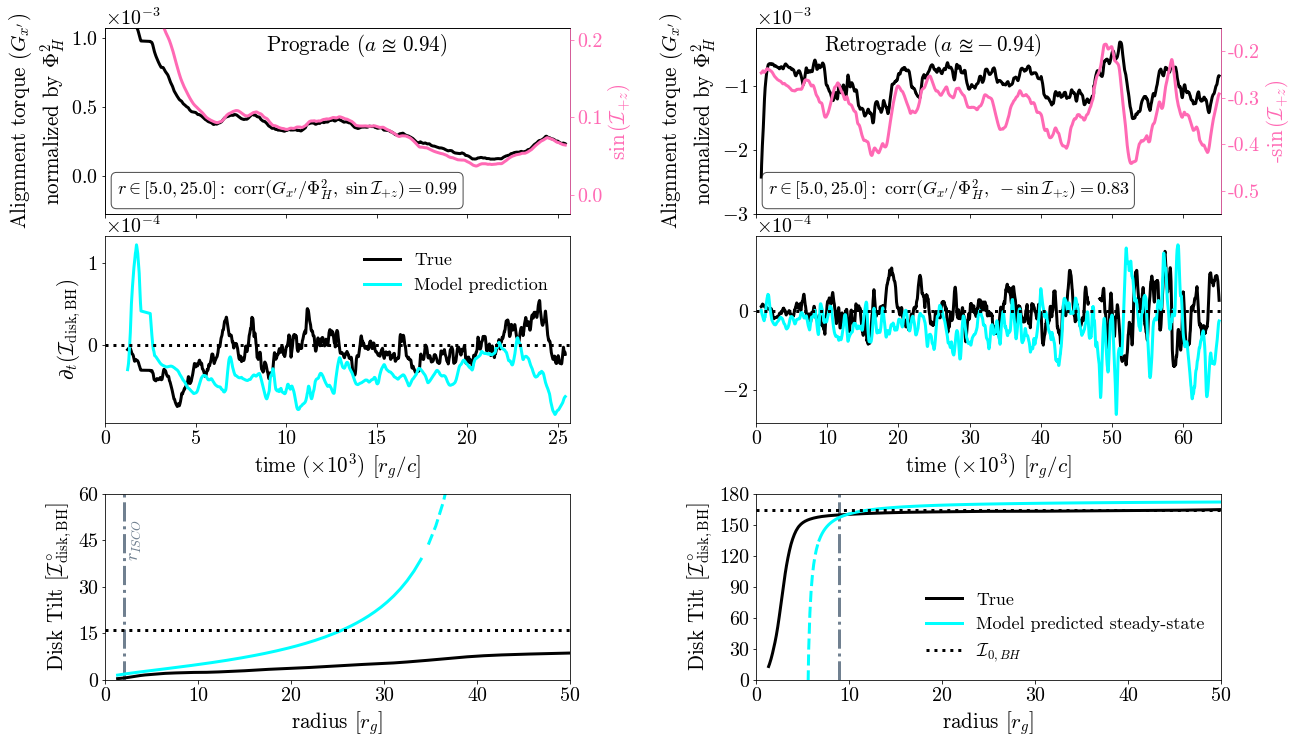}
\caption{\small{\textbf{Top row:} Time series of the normalized alignment torque, $G_{x^\prime}/\Phi_H^2$ (black, left $y$-axis), radially integrated over $r\in[5,25]\,r_g$, shown with $\sin\mathcal{I}_{+z}$ (pink, right $y$-axis) for the high-spin prograde run (left column) and $-\sin{\mathcal{I}_{+z}}$ for the high-spin retrograde run (right column). Annotated numbers are Pearson correlation, suggesting $G_{x^\prime} \propto\Phi_H^2 \sin{\mathcal{I}}_{+z}$ under a common sign convention. \textbf{Middle row:} Measured disk-tilt rate (black) compared to the empirical model (cyan; equation~\ref{eq:alignmentrarate_model}); the model reproduces the correct sign in the prograde run and captures episodic behavior in the retrograde run, but overpredicts the net alignment rate in both cases. \textbf{Bottom row:} Comparison of the late-time radial tilt profile, $\mathcal{I}_{\mathrm{disk},\mathrm{BH}}(r)$ (black; averaged over the final $10^4\,r_g/c$), with the model steady-state prediction (cyan; equation~\ref{eq:modelsteadystate}). The model captures the radial trend well, but diverges as the solution approaches its singularity (dashed segments) }}
\label{fig:empirical_model}
\end{figure*}

To test this hypothesis, we investigate the patterns in alignment torque ($G_{x^\prime}$) and shear term ($S_{x^\prime}$) across our high-resolution runs, and ask whether their variability can be factorized into global properties of disk, such as accretion rate, tilt, magnetic pressure and/or flux. Although, there is no particular reason for a simple scaling, for small tilts considered here, the projected stress components may, to first order, scale with magnetic pressure/flux. This motivates an empirical model for alignment torque. In the top row of Figure \ref{fig:empirical_model}, we plot time series of $G_{x^\prime}/\Phi_H^2$ (black curve) together with instantaneous tilt (pink curve) on a twin axis for high-spin prograde and retrograde flows in $r \in [5,25]\,r_g$. For prograde flows, we use $\sin{\mathcal{I}}_{+z}$, and for retrograde we use $-\sin{\mathcal{I}}_{+z}$ so that the expected handedness is the same and positive correlation implies a common driver. In both spins, the curves closely track each other; the annotated correlations are Pearson coefficients, indicating that $G_{x^\prime} \propto\Phi_H^2 \sin{\mathcal{I}}_{+z}$. We find this scaling to hold well for $r < 30\,r_g$.

To isolate the radial envelope, we time-averaged the ratio $\left<G_{x^\prime}\right>_t/\left<\Phi_H^2 \sin{\mathcal{I}}_{+z}\right>_t$ over several time-windows and found a simple power law $f(r) \sim A r^p$ fits all windows with minimal drift in $A$ and $p$.
This suggest a factorized description of alignment torque, i.e. $G_{x^\prime} \sim f(r) \Phi_H^2 \sin{\mathcal{I}}_{+z}$. Furthermore, performing the same fitting for other high-res runs revealed that $A \sim k_1 \Omega_H$, where $\Omega_H$ is angular frequency of the event horizon. The fitted parameters: $k_1 \approx 2.7 \times 10^{-2}$ and $p$ lies in the range $-0.25$ to $-0.3$ for prograde runs, and $-0.55$ for the high-spin retrpgrade case. The $30^\circ$ inclined case for $a = 0.5$ shows much steeper scaling with $p \sim -0.38$. Taken together, these trends produced a simple model of alignment torque: $G_{x^\prime} \sim k_1 r^p (\Omega_H \Phi_H^2) \sin{\mathcal{I}}_{+z}$ .

Next, we parametrize the shear term $S_{x^\prime}$ (equation \ref{eq:appendix_sheartermxprime}). We find that $\mathcal{S}_{x^\prime}(r,t) \approx -\frac{\partial \mathcal{I}_{+z}}{\partial r} G_{z^\prime}$ exceptionally well for prograde flows as disk aligns rapidly leaving accretion torque term the dominant force. Whereas, for retrograde, $-\partial_r(\mathcal{I}_{+z}) G_{z^\prime}$ is still dominant (on average, $75\%$ of shear-term comes from it), but in early times, the precession term becomes comparable as the precession profile sets early on in the disk evolution. 
So, keeping shear-term to first-order, we approximated $\mathcal{S}_{x^\prime}(r,t) \sim -\frac{\partial \mathcal{I}_{+z}}{\partial r} G_{z^\prime}$ for both cases, where $G_{z^\prime} = \left[-T^r_{~\phi^\prime}\right]_{S^\prime_r} \equiv -\dot{J}(r,t)$. Analyzing $\dot{J}$ uncovered that $\dot{J} \sim k_2 \Omega_H \Phi_H^2(t)$ with $k_2 \approx 1.1 \times 10^{-2}$ and only weak radial variation within $r < 30\,r_g$. This scaling resembles the Blandford-Znajek mechanism \citep{Blandford_Znajek_1977,Tchekhovskoy_2010} but with lower apparent efficiency because angular momentum flux includes both EM and hydrodynamic contributions. 

Combining these observations led to a minimal, data–driven model model for the alignment rate :
\begin{equation}
\label{eq:alignmentrarate_model}
\left[\partial_t(\mathcal{I}_{+z})\right]_{\text{model}}
  \sim
  \frac{\Omega_H \Phi_H^2}{|\boldsymbol L|_{S_r}}
  \left[
    \frac{\partial\bigl(k_1 r^p \sin{\mathcal{I}_{+z}}\bigr)}{\partial r} -
    k_2 \frac{\partial \mathcal{I}_{+z}}{\partial r}\right]
\end{equation}

We confront our empirical model to the simulation data in the middle panel of Figure \ref{fig:empirical_model} for radial region $r \in [15,25]\,r_g$. The model combines the fitted $f(r)$ with the measured $\Phi_H^2(t)$ and the disk tilt relative to the spin axis for both alignment torque and a shear-coupling correction. Without additional tuning in time, the model reproduces the correct sign of the disk tilt rate for the high-spin prograde case, and the muted, intermittent behavior in the high-spin retrograde case. However, for both cases it overestimates the alignment rate, indicating that it would not reproduce the disk tilt evolution. 

Finally, setting $\left[\partial_t(\mathcal{I}_{+z})\right]_{\text{model}} = 0$ gives the following differential equation, which under small-tilt approximation yields: 
\begin{align}
\label{eq:modelsteadystate}
    k_1 \frac{d (r^p \sin{\mathcal{I}_{+z})}}{dr} &= k_2 \frac{d \mathcal{I}_{+z}}{dr} \nonumber \\
    \mathcal{I}_{+z}(r) &\sim \frac{C}{k_1 r^p -k_2}
\end{align}
where the final step uses small-tilt approximation and $C$ is an integration constant. For our fitted parameters: $k_1 \approx 2.7 \times 10^{-2}$ and $k_2 \approx 1.1 \times 10^{-2}$, the above solution has a singular point at $r_{s} = (k_2/k_1)^{1/p} ~> 0$, where the solution changes behavior. For instance, if $\left(k_1 r^p - k_2\right) > 0$ and $r < r_s$, the disk tilt (relative to $+z$ axis) increases with radius, and decreases for $r > r_f$. Whereas, for $\left(k_1 r^p - k_2\right) < 0$ the trend changes. Furthermore, the singular point moves outward as $p$ becomes less negative. For instance, for $a = 0.75$ and $0.9375$ with $p \sim -0.32$ and $-0.24$, we find $r_s \sim 16\,r_g$ and $42\,r_g$, whereas for retrograde, we noted $r_s \sim 5\,r_g$. 

We compare the steady-state profiles of disk tilt for our high-spin prograde and retrograde cases against the steady-state profile predicted by the model in the bottom row of Figure \ref{fig:empirical_model}. We calculated $C = (k_1 r_{ref}^p -k_2)~\mathcal{I}_{+z}\big|_{r_{ref}}$, where we fixed $r_{ref} =10\,r_g$ as the system reached the viscous equilibrium there and our derived torques are valid only for $r \geq 5\,r_g$. The plot shows that the model captures the correct radial trend in both cases, especially for the retrograde run. The cause for it lies in the fact that $k_1 r^p - k_2 < 0$ for $r > r_s$ in retrograde case, producing decreasing $\mathcal{I}_{+z}$ with radius, and hence increasing radial profile of disk tilt relative to the spin axis. Close to the singular radius, the model prediction diverges as expected. 

Even though model has its limitations, we can still address the question of two-stage prograde alignment. In the pre-MAD era during which \emph{radial profile of tilt is nearly constant}, the alignment rate for a fixed spin is: $\partial_t(\mathcal{I}_{BH})_{\text{rapid}} \propto p k_1 \Phi_H^2 \sin{\mathcal{I}_{BH}} r^{p-1}$, i.e relatively stronger as the magnetic flux advects onto the hole. However, after the disk becomes MAD, $\Phi_H^2 \sin{\mathcal{I}_{BH}}$ decreases substantially leading to reduction in alignment rate. Furthermore the faster alignment at smaller radius produces $\partial_r(\mathcal{I}_{disk,BH}) > 0$, leaving $\partial_r(G_{x^\prime}) > 0$ in post-MAD era. The slow yet non-zero alignment rate comes from the shear-coupling term: $-k_2 \Omega_H \Phi_H^2 \partial_r(\mathcal{I}_{BH})$, which loosely quantifies the efficiency of accretion torque in reorienting the disk as it redistributes angular momentum across a warp.

\begin{figure*}[t!]
\centering
\includegraphics[width=\textwidth]{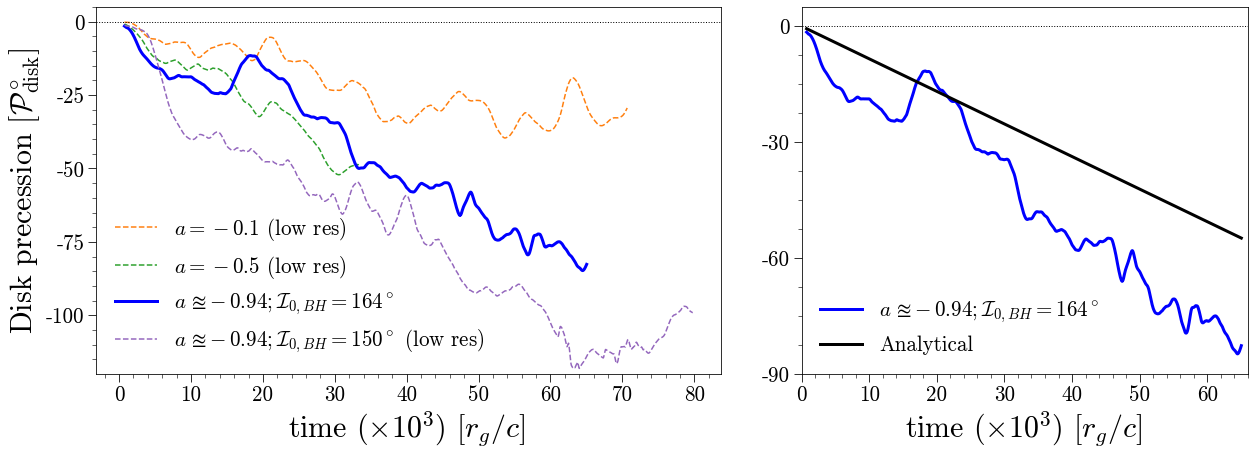}
\caption{\small{ \textbf{Left:} Disk precession angle, $\mathcal{P}_{\rm disk}(t)$ (equation~\ref{eq:diskprecessioneq}) integrated over $r\in[10,100]\,r_g$, for retrograde MADs with different spins and initial tilts (legend). The high-resolution $a\simeq-0.94$, $\mathcal{I}_{0,\rm BH}=164^\circ$ run is shown as the solid blue curve; dashed curves denote lower-resolution runs. Negative $\mathcal{P}_{\rm disk}$ corresponds to clockwise precession, i.e., in the same sense as the BH rotation. \textbf{Right:} High-spin retrograde case ($a\simeq-0.94$, $\mathcal{I}_{0,\rm BH}=164^\circ$; blue) compared to the rigid-body precession prediction of \cite{Fragile_2007} (black; evaluated for $r_i\simeq10\,r_g$, $r_o\simeq100\,r_g$, and $\zeta\simeq-0.5$). The simulated precession is faster, with a period $\sim2$--$5\times$ shorter than the analytic estimate. }} 
\label{fig:retrograde_MADs_disk_precession}
\end{figure*}

\section{Solid-Body Precession of the Retrograde MADs}
\label{sec:Solid Body Precession of the Retrograde MADs}
Building on our finding that retrograde MADs do not align, we measured the precession of the disk about the fixed $+ z$-axis, following the definition from \cite{Sajal2024}:
\begin{equation}
\label{eq:diskprecessioneq}
    \mathcal{P}_{\text{disk}}(r,t) = \tan^{-1}\left(\frac{\langle L_{y,\text{gas}} \rangle_{\rho}}{\langle L_{x,\text{gas}} \rangle_{\rho}}\right)
\end{equation}
The left panel of Figure \ref{fig:retrograde_MADs_disk_precession} shows the total disk precession angle for each retrograde MAD run in the region $10\,r_g \leq r \leq 100\,r_g$ region. The high resolution $a \approxeq -0.94$ is shown as a solid, thick blue line, while the low-resolution cases appear as thin, dashed colored curves. The negative angles indicating clockwise motion in the same direction as the BH's rotation. We find that precession rate grows with increasing spin magnitude, and in the high-spin retrograde run doubling the initial tilt from $16^\circ$ to $30^\circ$ (relative to $+ z$-axis) increases the precession frequency by roughly $50\%$. Although the low-resolution run hint at a tilt dependence, their coarse resolution may also amplify the precession rate \citep{Liska2018jetprecess,Chatterjee2020}.

We compare our observations to the analytic model of \cite{Fragile_2007} in the right panel of Figure \ref{fig:retrograde_MADs_disk_precession}, assuming inner and outer disk radius of $r_i \sim 10\,r_g$ and $r_0 \sim 100\,r_g$, and surface density slope $\zeta \sim -0.5$, derived from the surface density profile $\Sigma = \Sigma_i (r/r_i)^{-\zeta}$ for the high-spin retrograde run with tilt $= 164^\circ$. We find that the observed precession period is consistently 2–5 times shorter than the analytical estimate across all simulated spins, suggesting a faster precession rate than the frequency of a test-particle at the radius of the disk’s mean angular momentum. 

\section{Discussion} \label{sec:discussion}
\subsection{Can the Retrograde MADs Align?}
Our results demonstrate that high-spin retrograde MADs exhibit cyclical alignment tendencies, yet gas inertia ultimately counteracts magnetic forces, preventing complete alignment (Figure \ref{fig:density_plot_and_flux_and_disk_inclination_comparison} and \ref{fig:torque_contribution_to_rate_of_change_of_tilt}). This raises a critical question: under what conditions, if any, can retrograde MADs achieve mutual alignment with the BH spin? To discuss it, we re-write equation \ref{eq:appendix_didt_indiskframetorques} for \emph{retrograde tilts} as:
\begin{equation}
    |\boldsymbol{L}|_{S_r}\partial_t(\mathcal{I}_{BH}) = -\partial_r(\tau_{x^\prime}) + \mathcal{S}_{x^\prime}
\end{equation}
The above equation dictates that for retrograde flows to align the net action of alignment force and the shear-term should be negative. Assuming our empirical model holds at larger tilts and horizon magnetic flux remain constant, it suggest that the alignment torque grows with misalignment and decreases with radius. Thus, $-\partial_r(\tau_{x^\prime})$ should be more negative at larger tilts and BH spin magnitude. 

For the shear-term, which is positive in our retrograde models and thus opposes alignment, we therefore expect the approximation $\mathcal{S}_{x^\prime} \approx \partial_r(\mathcal{I}_{BH}) G_{z^\prime}$ (for retrograde) to remain relevant at larger tilts. First, increasing tilt reduces $\cos{\mathcal{I}_{BH}}$, so the second term contribution $\propto -G_{y^\prime}^{\text{MA+EM}} \cos{\mathcal{I}_{BH}} \partial_r(\mathcal{P})$ in shear-term (equation \ref{eq:appendix_sheartermxprime}) is naturally suppressed. Second, if radial twist profile is set by LT torque, then one may estimate $\partial_r(\mathcal{P}) \sim \Omega_{LT}/|v^r|$, with $\Omega_{LT} \propto r^{-3}$ and $v_r \sim r^{-1}$ in MAD simulations \citep{Narayan2022}, so $\partial_r(\mathcal{P}) \propto r^{-2}$ and hence, the combination $\cos{\mathcal{I}_{BH}} \partial_r(\mathcal{P})$ is expected to decrease with increasing tilt at fixed radius. Finally, local maxwell stress associated with accretion is expected to dominate relative to precessional torque: $G_{y^\prime}$ which is arising because of stress projections, indicating $G_{z^\prime} > G_{y^\prime}$. Under these assumptions the shear term should effectively scales as $\partial_r(\mathcal{I}_{BH}) G_{z^\prime}$. Moreover, recent simulations of \cite{Chatterjee2023misaligned} indicate that jet power (and thus the associated EM torque) declines with increasing tilt, implying $G_{z^\prime}$ is likely reduced at larger misalignment, further weakening this opposing geometric term.

Combining these trends suggests that there may exist a critical tilt angle beyond which the growth of the alignment torque with misalignment can outweigh the weakening shear term in the inner MAD region. In such a regime the retrograde flow could, in principle, drive toward a mutually aligned state. Conversely, if the alignment torque saturates or scales only a $\sin{\mathcal{I}_{BH}} \cos{\mathcal{I}_{BH}}$, its growth with misalignment would be too weak and hot retrograde MADs would remain persistently misaligned. Distinguishing between these possibilities requires simulations that explicitly measure the tilt dependence of the alignment torque (see section \ref{subsec:caveats}).

\subsection{Precession of Prograde Flows and the Jet}
Relativistic jets in retrograde flows, which do not emit along the BH rotation axis, may undergo precession \citep{Liska2018jetprecess,Chatterjee2023misaligned}. For the weakly rotating BH ($a = -0.1$), we observe precession of the jet occurring at a rate comparable to the disk. At higher spins, however, numerical fluctuations in the jet's twist angle\textemdash driven by inflowing gas disrupting magnetized regions ($\sigma > 1$)\textemdash preclude robust measurement of precession period. Nevertheless, the observed $a = -0.1$ behavior and our setup’s similarity to \cite{Chatterjee2023misaligned} suggest jet precession remains plausible in astrophysical systems.

While investigating precession in prograde flows, we observed an unexpected pattern. Although near-complete alignment during flux saturation should render precession moot, tracking the twist angle over time reveals prograde precession, followed by a transition to clockwise, retrograde, solid-body precession in slow-alignment phase. The same phenomenon is also observed by \cite{Jiang_2025}. To validate this, we ran a low resolution, high-spin BH ($a \approxeq -0.94$) with a $164^\circ$-tilted torus, effectively simulating a prograde case. Here, too, the disk precess opposite to the BH spin. Analyzing precession torque gradient $\partial_r(G_{y^\prime})$ (as $\mathcal{S}_{y^\prime} \ll \partial_r(G_{y^\prime})$ in equation \ref{eq:appendix_dtwistdt_indiskframetorques} for $r \geq 10\,r_g$) revealed that inertial torques dominate as gravitational torques weaken during rapid alignment, driving this counter-rotating precession. However, more in-depth analysis is needed to conclude it. 

\subsection{Comparison to Previous Works}
Our results broadly agree with prior GRMHD studies of misaligned tori \citep{Mckinney_tilted_2013,Chatterjee2023misaligned}, confirming that prograde MADs align post flux saturation, even for spins as low as $a = 0.3$, supporting the hypothesis from \cite{Mckinney_tilted_2013} that magneto-spin alignment may be achievable for most or all geometrically thick, prograde MAD spins. Like \cite{Mckinney_tilted_2013}, we find disk-jet alignment within $50\,r_g$ for prograde flows, with divergence at larger radii attributed to jet interactions with inflowing gas.

A key discrepancy arises for retrograde MADs: while \cite{Mckinney_tilted_2013} reported disk flipping and eventual alignment of it with the spin, our retrograde flows remain misaligned. This difference likely stems from how tilt is introduced. In \cite{Mckinney_tilted_2013}, the relative tilt between the disk (and magnetosphere) and the BH was imposed mid-simulation by rotating the spin axis in a previously aligned MAD. This preserved magnetic flux levels comparable to prograde cases, as noted in \cite[Table 9]{Mckinney_2012}, enabling strong alignment torques. In contrast, initial tilt introduction in our setup suppresses flux saturation in retrograde flows, weakening magnetic torques below the threshold for alignment.

Our torque decomposition further suggests that the hydrodynamic flux is not merely a residual term: it can respond to the magnetic alignment torque and thereby modulate the net alignment rate (Section~\ref{sec:Matter, Electromagnetic and Gravitational Torque}). Prior simulations show robust alignment in prograde MADs (except at very large tilts; \cite{Chatterjee2023misaligned}) but incomplete alignment and enhanced warping in more weakly magnetized SANE-like flows \citep{Fragile_2007,White_2019,Sajal2024}; this leaves open the possibility that a relatively narrow range of horizon-threading magnetic flux separates these regimes, with the transition potentially complicated by temporal flux variability \citep{White_2020_semiMAD}

Another difference concerns the physical interpretation of the alignment mechanism. In our simulations the jet and inner disk align on similar rate, and the jet axis remains nearly perpendicular to the disk plane, which makes it difficult for the jet alone to provide the lateral torque needed to reorient the disk. Instead, our torque decomposition indicates that the dominant contribution to $G_{x^\prime}$ which led the initial rapid alignment arises from the in–plane $T^r_{\phi^\prime}$ stress projected into the tilted basis, i.e. from non-axisymmetric magnetic correlations within the disk rather than from the axisymmetric component that primarily feeds the jet. Because these stresses may still scale with the horizon magnetic flux, the net alignment torque may superficially resembles a jet–driven picture, even if the underlying torques are disk-driven. We therefore interpret magneto-spin alignment in our models as being mediated mainly by internal disk stresses, while noting that a more quantitative separation of jet and disk contributions will require analysis over a wider range of tilts and magnetic strength.

\subsection{Validity of the Empirical Model}
\label{subsec:caveats} 
In section \ref{subsec:Empirical_model_two_stages_alignment}, we modeled alignment torque as: $G_{x^\prime} \propto \Phi_H^2 \sin{\mathcal{I}_{+z}}$. A related form $G_{x^\prime } \propto \Phi_H^2\sin{\mathcal{I}_{+z}} \cos{\mathcal{I}_{+z}}$ shows a comparably good correlation, except very close to the horizon in retrograde flows. Because our simulations only probe modest tilts, we cannot robustly distinguish between these functional forms, and the explicit tilt dependence of the alignment torque remains poorly constrained. An additional factor such as $\cos{\mathcal{I}_{+z}}$ could help explain explain why the extremely misaligned disks in \cite{Chatterjee2023misaligned} do not align. 

A further limitation is the degeneracy in how the torque scales with magnetic strength. While we have parameterized the alignment torque in terms of $\Phi_H$, similar if not better correlations are obtained with local magnetic flux or magnetic pressure. Since disk alignment ultimately depends on the local angular-momentum flux, it is plausible that scaling $G_{x^\prime}$ and $G_{y^\prime}$ with local magnetic pressure would provide a more accurate description at larger radii; we have retained $\Phi_H$ here because it displayed moderately better correlations in prograde flows. Finally, our empirical model is constructed for the net torque. It is possible that a simple flux-based scaling applies more cleanly to the EM component alone, particularly in tilted, weakly magnetized (SANE-like) flows.

\section{Conclusion} \label{sec:conclusion}
In this study, we set out to elucidate the alignment processes in tilted MADs around BHs of varying spin parameters. Using a set of global, ideal 3D GRMHD simulations we explored accretion scenarios with spins ranging from $a = -0.1$ to $a = 0.9375$, incorporating initial disk tilts of $16^\circ$ and/or $30^\circ$ relative to the $+z$-axis. The primary findings of our study are as follows:
\begin{enumerate} 
\item Tilted prograde MADs achieve mutual alignment with the BH spin axis through a two-stage process. The disk first undergoes rapid alignment on the magnetic flux saturation timescale, followed by a slower phase in which the inner disk continues to align at a reduced but measurable rate.
\item Retrograde MADs remain persistently misaligned and exhibited near-solid body precession, with periods $\sim10^5\,r_g/c$, consistent with the $\sim 0.2$\,Hz Type-C QPOs observed in $\sim 10\,M_{\odot}$ black hole X-ray binaries \citep{Ingram2016,Liska2018}. In both prograde and low-spin retrograde systems, jets emerge perpendicular to the inner disk midplane. 
\item EM torques are the primary drivers of disk alignment in both prograde and retrograde MADs\textemdash but with significantly greater strength in prograde flows. In retrograde cases, opposing hydrodynamic torques, effectively counterbalance the EM torque, preventing alignment, while gravitational torques remain negligible.
\item Evaluating the EM contribution to the alignment torque with narrow polar caps excluded shows that the (projected) radial–azimuthal magnetic stress is the primary driver of prograde alignment, while in retrograde flows all EM channels are substantially weaker.
\item Across the high-resolution runs, the net alignment torque per shell is well described by a separable scaling $G_{x^\prime} \propto f(r)\,\Omega_H \Phi_H^2 \sin\mathcal{I}_{+z}$ with a shallow $f(r)$, which preliminary explained the observed two-stage prograde alignment but our empirical model systematically overestimates the simulation-measured tilt rate.
\end{enumerate}

\begin{acknowledgments}
S.G. gratefully acknowledges stimulating discussions with M. C. Begelman, K. Long, C. Echiburú-Trujillo, P. Dhang, and J. Jacquemin Ide. This work was supported in part by NASA Astrophysics Theory Program grant 80NSSC24K1094 and Chandra awards TM3-24003X and TM4-25005X.
\end{acknowledgments}

\software{Matplotlib \citep{Matplotlib}, NumPy \citep{NumPy}}

\appendix
\section{Rate of Change of angular momentum}
\label{sec:Rate of Change of angular momentum}
The goal of this section is to provide a rationale for the use of total angular momentum to study the disk tilt and to elucidate the identification of torque components while deriving the evolution equations of the angular momentum.

\subsection{Tilt using total angular momentum}
\label{sec:Appendix_Tilt using total angular momentum}
\begin{figure*}[!ht]
\centering
\includegraphics[width=\textwidth]{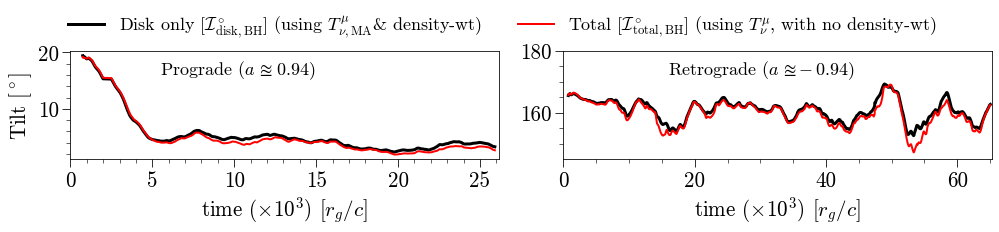}
\caption{\small{ ime evolution of the disk tilt in the inner region $r\in[3,20]\,r_g$, comparing the density-weighted tilt computed from $T^{\mu}{}_{\nu,\mathrm{MA}}$ (black; equations~\ref{eq:diskangmomcompeq} and~\ref{eq:diskinclinationeq}) to the total tilt computed from $T^{\mu}{}_{\nu}$ without density weighting (red; equations~\ref{eq:totalangmomcompeq} and~\ref{eq:totalinclinationeq}), for high-spin prograde (left) and retrograde (right) runs with initial tilt $16^\circ$ relative to $+z$. This comparison motivates using the total tilt when analyzing the disk angular-momentum evolution. }} 
\label{fig:appendix_comparing_tilt_of_disk_with_total}
\end{figure*}

Similar to equation \ref{eq:diskangmomcompeq}, we define the total angular momentum density, including both gas and magnetic fields, as:
\begin{equation}
\label{eq:totalangmomcompeq}
    (\mathbf{L_{\text{total}}})_{\hat{k}} = \epsilon_{ijk} r^{\hat{j}}~(T^t_{~\hat{k}}),
\end{equation}
where $T^{\mu}_{~\nu}$ is the total stress-energy tensor and the subscript ``total'' signifies the combined contribution. The total angular momentum vector serves as a reliable indicator of disk tilt for two main reasons. First, the magnetic field's inertia is negligible relative to the gas, making $(\mathbf{L_{\text{total}}})_{\hat{k}} \approx (\mathbf{L_{\text{gas}}})_{\hat{k}}$. Second, tilt and twist are angular properties expressed through the ratio of angular momentum components (see equations \ref{eq:diskinclinationeq} and \ref{eq:diskprecessioneq}), and thus are not influenced by the absolute magnitude of any single component. Figure \ref{fig:appendix_comparing_tilt_of_disk_with_total} validates the approximation by comparing disk tilt (from section \ref{sec:section of Alignment and Misalignment of Prograde and Retrograde MADs}) to tilt derived from total angular momentum (equation \ref{eq:totalangmomcompeq}) as:
\begin{equation}
\label{eq:totalinclinationeq}
    \mathcal{I}_{\text{total},\text{BH}}(r,t) \equiv \cos^{-1}\left(\frac{\mathbf{L}_{BH}\cdot\left[\mathbf{L_{\text{total}}}\right]_{S_r}}{|\mathbf{L}_{BH}||\left[\mathbf{L_{\text{total}}}\right]_{S_r}|}\right),
\end{equation}
where $[x]_{S_r} $ is the area integral (equation \ref{eq:surfaceintegral}). We did not use density-weighting in equation \ref{eq:totalangmomcompeq} to ensure the contributions from highly magnetized regions as well as the high-density areas are on same footing. While the comparison in Figure \ref{fig:appendix_comparing_tilt_of_disk_with_total} covers the radial range $[3, 20]\,r_g$, we found this approximation to hold true up to $300\,r_g$, the extent of the initial torus. However, close to the horizon, magnetic contributions rival gas inertia leading to the breakdown of the approximation.

\subsection{Torque equations}
\label{sec:Appendix_Torque equations}
Building on the rationale for using total angular momentum density to study disk tilt, we derive the torque equations governing total angular momentum evolution. \emph{For the following sections, we report every quantity relative to $+z$ axis of the simulation frame.} We begin by expressing the total angular momentum components in Cartesian coordinates, where the stress-energy tensor components are described in spherical coordinates. The total angular momentum in a spherical region of radius `$r$' at a fixed time is given by:
\begin{equation}
\label{eq:appendix_angularmomcompsvolintegrated}
    \begin{split}
    \int_{r_{i}}^{r} L_{x,\text{total}}~dV &= \int_{r_{i}}^{r} \left(-\sin{\phi}~T^t_{\theta} - \cot{\theta}\cos{\phi}~T^t_{\phi}\right) dV\\
    \int_{r_{i}}^{r} L_{y,\text{total}}~dV &= \int_{r_{i}}^{r} \left(\cos{\phi}~T^t_{\theta} - \cot{\theta}\sin{\phi}~T^t_{\phi} \right) dV, \\
    \int_{r_{i}}^{r} L_{z,\text{total}}~dV &= \int_{r_{i}}^{r} T^t_{\phi}~dV,
\end{split}
\end{equation}
where $dV=\sqrt{-g}\,dr\,d\theta\,d\phi$ is the proper volume element. To obtain the time evolution of the angular momentum in this control volume, we differentiate equation~\ref{eq:appendix_angularmomcompsvolintegrated} and use energy--momentum conservation (\ref{eq:GRMHD2}) to express the rate in flux form as
\begin{equation}
\label{eq:appendix_fluxform_identity}
\partial_t\!\left(\sqrt{-g}\,T^t{}_\nu\right)
= -\,\partial_i\!\left(\sqrt{-g}\,T^i{}_\nu\right)
+ \sqrt{-g}\,T^\kappa{}_\lambda\,\Gamma^\lambda_{\nu\kappa},
\end{equation}

For a stationary, axisymmetric metric we have $\partial_t\sqrt{-g}=0$, and the
azimuthal Killing symmetry implies $\sqrt{-g}\,T^\kappa{}_\lambda\Gamma^\lambda_{\phi\kappa}=0$. Substituting equation~\ref{eq:appendix_fluxform_identity} into the definitions of $L_i$ and using the product rule to absorb the trigonometric prefactors into the divergence terms, we obtain

\begin{equation}
\label{eq:appendix_Lxvolintegratedtimederivativefluxformandreaarangement}
\begin{split}
    \partial_t \left(\int_{r_{i}}^{r} L_{x,\text{total}}~dV \right) &= \int_{r_{i}}^{r} \partial_i (\sqrt{-g} \sin{\phi}~T^i_{\theta}) dr d\theta d\phi~+ \int_{r_{i}}^{r} \partial_i (\sqrt{-g} \cot{\theta} \cos{\phi}~T^i_{\phi}) dr d\theta d\phi~+ \\
    & -\int_{r_{i}}^{r} \sin{\phi}~T^\kappa_\lambda \Gamma^\lambda_{\theta \kappa}~dV ~+\int_{r_{i}}^{r} \left(\cot{\theta}\sin{\phi}~T^\phi_{\phi} + \frac{\cos{\phi}}{\sin^2{\theta}} T^\theta_{\phi} - \cos{\phi}~T^\phi_{\theta}\right) dV
\end{split}
\end{equation}

\begin{equation}
\label{eq:appendix_Lyvolintegratedtimederivativefluxformandreaarangement}
\begin{split}
    \partial_t \left(\int_{r_{i}}^{r} L_{y,\text{total}}~dV \right) &= -\int_{r_{i}}^{r} \partial_i (\sqrt{-g} \cos{\phi}~T^i_{\theta}) dr d\theta d\phi~+ \int_{r_{i}}^{r} \partial_i (\sqrt{-g} \cot{\theta} \sin{\phi}~T^i_{\phi}) dr d\theta d\phi~+ \\
    & \int_{r_{i}}^{r} \cos{\phi}~T^\kappa_\lambda \Gamma^\lambda_{\theta \kappa}~dV ~+\int_{r_{i}}^{r} \left(-\cot{\theta}\cos{\phi}~T^\phi_{\phi} + \frac{\sin{\phi}}{\sin^2{\theta}} T^\theta_{\phi} - \sin{\phi}~T^\phi_{\theta}\right) dV
\end{split}
\end{equation}

\begin{equation}
\label{eq:appendix_Lzvolintegratedtimederivativefluxform}
    \partial_t \left(\int_{r_{i}}^{r} L_{z,\text{total}}~dV \right) = \int_{r_{i}}^{r} \left[\partial_i(-\sqrt{-g}T^i_{\phi})\right]~dr d\theta d\phi
\end{equation}

We now focus on each component individually, beginning with the axisymmetric `$z$' component. The RHS of equation \ref{eq:appendix_Lzvolintegratedtimederivativefluxform} represents the net flux of azimuthal momentum out of the control volume at a given time. Since we study the local rate of change of total angular momentum in the Eulerian frame and in the 3+1 decomposition of spacetime, the control volume, fixed in space, corresponds to a hypersurface of constant time. This allows us to rewrite the RHS in terms of proper volume integration. Applying the Gauss-divergence theorem \citep{MTWGravitation,Bertschinger1999,Carrollbook}, we obtain:
\begin{equation}
\label{eq:appendix_Lzgaussdivfluxform}
\begin{split}
    \partial_t \left(\int_{r_{i}}^{r} L_{z,\text{total}}~dV \right) &= \int_{r_{i}}^{r} \left[\partial_i(-\sqrt{-g}T^i_{\phi})\right]~dr d\theta d\phi = -\int_{r_{i}}^{r} \nabla_i(\sqrt{-g}T^i_{\phi})~dV = \left(-\int T^r_{\phi}~dS_r\right)\Biggr|_{r_i}^{r}
\end{split}
\end{equation}

Performing the same set of operations for equations \ref{eq:appendix_Lxvolintegratedtimederivativefluxformandreaarangement} and \ref{eq:appendix_Lyvolintegratedtimederivativefluxformandreaarangement} gives:

\begin{equation}
\label{eq:appendix_Lxgaussdivfluxform}
\begin{split}
    \partial_t \left(\int_{r_{i}}^{r} L_{x,\text{total}}~dV \right) &= \left(\int \left(\sin{\phi}~T^r_{\theta} + \cot{\theta} \cos{\phi}~T^r_{\phi}\right) dS_r \right)\Biggr|_{r_i}^{r} -\int_{r_{i}}^{r} \sin{\phi}~T^\kappa_\lambda \Gamma^\lambda_{\theta \kappa}~dV ~+ \\
    &\int_{r_{i}}^{r} \left(\cot{\theta}\sin{\phi}~T^\phi_{\phi} + \frac{\cos{\phi}}{\sin^2{\theta}} T^\theta_{\phi} - \cos{\phi}~T^\phi_{\theta}\right) dV
\end{split}
\end{equation}

\begin{equation}
\label{eq:appendix_Lygaussdivfluxform}
\begin{split}
    \partial_t \left(\int_{r_{i}}^{r} L_{y,\text{total}}~dV \right) &= \left(\int \left(-\cos{\phi}~T^r_{\theta} + \cot{\theta} \sin{\phi}~T^r_{\phi}\right) dS_r \right)\Biggr|_{r_i}^{r} ~+ \int_{r_{i}}^{r} \cos{\phi}~T^\kappa_\lambda \Gamma^\lambda_{\theta \kappa}~dV ~+ \\
    &\int_{r_{i}}^{r} \left(-\cot{\theta}\cos{\phi}~T^\phi_{\phi} + \frac{\sin{\phi}}{\sin^2{\theta}} T^\theta_{\phi} - \sin{\phi}~T^\phi_{\theta}\right) dV
\end{split}
\end{equation}

The equations above give the time rate of change of the total angular-momentum components enclosed within a closed spherical-shell volume between $r_i$ and $r$. To study the radial structure and time-averaged profiles of disk tilt and twist, we require the corresponding \emph{local} evolution at each radius. Following \cite{Sorathia_2014}, we take a radial derivative and apply the Leibniz rule to convert the volume integrals into surface terms evaluated at $r$. We then decompose $T^\mu{}_\nu$ into matter and electromagnetic parts \citep{Gammie_2003} to isolate the contributions from different torque channels, yielding the expressions below.
\begin{equation}
\label{eq:appendix_Lztorquefinal}
\begin{split}
    \partial_t \left(\int L_{z,\text{total}}~dS_r \right) &= \partial_r\left(G_z\right) \equiv \underbrace{-\partial_r \left(\int \left(L_{z,\text{gas}} V^r\right) dS_r \right)}_{\text{Matter torque (z-comp)}} ~ \underbrace{- \partial_r \left( \int T^r_{\phi,\text{EM}} ~dS_r\right)}_{\text{EM torque (z-comp)}},
\end{split}
\end{equation}

\begin{equation}
\label{eq:appendix_Lxtorquefinal}
\begin{split}
    \partial_t \left(\int L_{x,\text{total}}~dS_r \right) &= \partial_r\left(G_x\right) \equiv \underbrace{-\partial_r \left(\int \left(L_{x,\text{gas}} V^r\right) dS_r \right)}_{\text{Matter torque (x-comp)}} ~+ \underbrace{\partial_r \left( \int\left(\sin{\phi}~T^r_{\theta,\text{EM}} + \cot{\theta} \cos{\phi}~T^r_{\phi,\text{EM}}\right)~dS_r\right)}_{\text{EM torque (x-comp)}} ~+ \\ 
    &\underbrace{\int \left(-\sin{\phi}~T^\kappa_\lambda \Gamma^\lambda_{\theta \kappa} + \cot{\theta}\sin{\phi}~T^\phi_{\phi} + \frac{\cos{\phi}}{\sin^2{\theta}} T^\theta_{\phi} - \cos{\phi}~T^\phi_{\theta}\right) dS_r}_{\text{Gravitational torque (x-comp)}},
\end{split}
\end{equation}

\begin{equation}
\label{eq:appendix_Lytorquefinal}
\begin{split}
    \partial_t \left(\int L_{y,\text{total}}~dS_r \right) &= \partial_r\left(G_y\right) \equiv \underbrace{-\partial_r \left(\int \left(L_{y,\text{gas}} V^r\right) dS_r \right)}_{\text{Matter torque (y-comp)}} ~+ \underbrace{\partial_r \left( \int \left(-\cos{\phi}~T^r_{\theta,\text{EM}} + \cot{\theta} \sin{\phi}~T^r_{\phi,\text{EM}}\right)~dS_r\right)}_{\text{EM torque (y-comp)}} ~+ \\ 
    &\underbrace{\int \left(\cos{\phi}~T^\kappa_\lambda \Gamma^\lambda_{\theta \kappa} - \cot{\theta}\cos{\phi}~T^\phi_{\phi} + \frac{\sin{\phi}}{\sin^2{\theta}} T^\theta_{\phi} - \sin{\phi}~T^\phi_{\theta}\right) dS_r}_{\text{Gravitational torque (y-comp)}},
\end{split}
\end{equation}
where $V^r \equiv u^r/u^t$ is the radial coordinate three-velocity, and $L_{i,\mathrm{gas}}$ is the gas-only angular-momentum density (equation~\ref{eq:diskangmomcompeq}), i.e., the matter contribution to the volume integral in equation~\ref{eq:appendix_angularmomcompsvolintegrated}. The quantities $G_i$ denote the components of the total torque vector (gas plus fields) in the simulation frame. This decomposition shows that the matter torque is set by the radial flux of gas angular momentum, while the electromagnetic torque arises from the divergence of the Maxwell stress, consistent with \cite{Sorathia_2014}. We define the total angular momentum magnitude as as $\left|\left[\boldsymbol L_{\rm total}\right]_{S_r}\right| \equiv \sqrt{[L_{x,\mathrm{total}}]^2_{S_r}+[L_{y,\mathrm{total}}]^2_{S_r} + [L_{z,\mathrm{total}}]^2_{S_r}}$. The next section describes our procedure for isolating the gravitational torque.

\subsection{Identification of Gravitational torque}
We identify the terms involving the metric connection ($T^\kappa{}_\lambda\Gamma^\lambda_{\theta\kappa}$) and the mixed $\theta\phi$ and $\phi\phi$ stresses in equations~\ref{eq:appendix_Lxtorquefinal}--\ref{eq:appendix_Lytorquefinal} as the gravitational-torque contribution. Below we validate this identification by showing that it reduces to the expected Lense-Thirring (LT) torque \citep{Lense-Thirring} in the weak-field limit.

We first examine the connection term $T^\kappa{}_\lambda\Gamma^\lambda_{\theta\kappa}$, which encodes latitudinal forces produced by curved spacetime, centrifugal and magnetic hoop stress, as well as other complex interactions that are not straightforward to interpret \citep{Scepi2023}. While $\Gamma^\lambda_{\theta\kappa}$ can be written in terms of metric gradients, its contraction with $T^\kappa{}_\lambda$ makes the dominant contribution non-obvious. Guided by several simulations (this work, \citealt{Sajal2024}, and the aligned high-spin MAD of \citealt{Dexter2020b}), we find that for $r\gg a$ the contraction is well-approximated by two toroidal-stress terms:
\begin{equation}
    \label{eq:Christoffelapproximation}
    T^\kappa_\lambda \Gamma^\lambda_{\theta \kappa} \approx T^{\phi}_{\phi} \Gamma^\phi_{\theta \phi} + T^t_{\phi} \Gamma^\phi_{\theta t}
\end{equation}
This behavior is consistent with the predominantly azimuthal motion of a weakly tilted flow and the buildup of toroidal magnetic stress, though we do not attempt a first-principles derivation. We next test whether our ``gravitational'' terms reproduce the LT form in the weak-field regime, i.e. $\mathbf{\Omega_{LT}} \times \mathbf{L_{\text{total}}} = \{-\Omega_{LT} L_{y,\text{total}},\Omega_{LT} L_{x,\text{total}},0\}$ with
$|\boldsymbol{\Omega}_{\rm LT}|=2a/r^3$ aligned with the BH spin axis

To approximating $T^{\theta}_{\phi}$ and $T^\kappa_\lambda \Gamma^\lambda_{\theta \kappa}$, we write the line element in the weak-field approximation ($a \ll 1$) \citep{Lense-Thirring,adler1965introduction}:
\begin{equation}
    ds^2 = -\left(1 - \frac{2}{r}\right)dt^2 + \left(1 + \frac{2}{r}\right)dr^2 + r^2\left(d\theta^2 + \sin^2{\theta}d\phi^2\right) - \frac{4 a \sin^2{\theta}}{r} d\phi dt ,
\end{equation}
and use tensor transformation rules to re-write equation \ref{eq:Christoffelapproximation} as:
\begin{equation}
\label{eq:extratermsandchristoffelapproximate}
    T^{\theta}_{\phi} \approx -\frac{2a \sin^2{\theta}}{r^3} T^t_{\theta} + \sin^2{\theta}~T^{\phi}_{\theta}; \quad T^\kappa_\lambda \Gamma^\lambda_{\theta \kappa} \approx T^{\phi}_{\phi} \Gamma^\phi_{\theta \phi} + T^t_{\phi} \Gamma^\phi_{\theta t} \approx \cot{\theta}~T^\phi_{\phi} - \frac{2 a \cot{\theta}}{r^3} T^t_{\phi} ,
\end{equation}
valid for $a\ll1$ and $r\gg a$. Substituting above equation into gravitational torque expressions (equation \ref{eq:appendix_Lxtorquefinal} and \ref{eq:appendix_Lytorquefinal}) yields:
\begin{equation}
\label{eq:appendix_Lxtorquegravity}
    \partial_r(G_x^{\text{gravity}}) \approx \int \frac{2a}{r^3}\left( -\cos{\phi}~T^t_{\theta} + \cot{\theta}\sin{\phi} T^t_{\phi} \right) dS_r \approx \int_\theta \int_\phi \left(- |\mathbf{\Omega_{LT}}| L_{y,\text{total}} \right) dS_r
\end{equation}
\begin{equation}
\label{eq:appendix_Lytorquegravity}
    \partial_r(G_y^{\text{gravity}}) \approx \int \frac{2a}{r^3}\left( -\sin{\phi}~T^t_{\theta} - \cot{\theta}\cos{\phi} T^t_{\phi} \right) dS_r \approx \int_\theta \int_\phi \left(|\mathbf{\Omega_{LT}}| L_{x,\text{total}} \right) dS_r
\end{equation}

where the final step uses equations~\ref{eq:appendix_angularmomcompsvolintegrated} and~\ref{eq:extratermsandchristoffelapproximate}. Thus our identification recovers the LT torque to leading order. Moreover, combinations such as $\frac{\cos{\phi}}{\sin^2{\theta}} T^\theta_{\phi} - \cos{\phi}~T^\phi_{\theta}$, and $\frac{\sin{\phi}}{\sin^2{\theta}} T^\theta_{\phi} - \sin{\phi}~T^\phi_{\theta}$ vanish in Schwarzschild and appear only in rotating spacetimes, consistent with their interpretation as frame-dragging corrections.

\begin{figure*}[!ht]
\centering
\includegraphics[width=\textwidth]{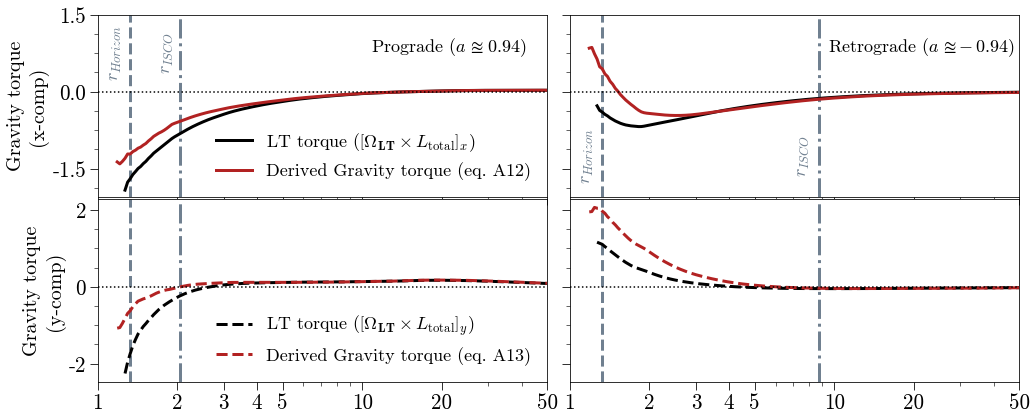}
\caption{\small{Radial profiles of the gravitational torque derived in Appendix~\ref{sec:Appendix_Torque equations} (red; equations~\ref{eq:appendix_Lxtorquefinal} and~\ref{eq:appendix_Lytorquefinal}) compared to the LT torque (black) for high-spin prograde (left) and retrograde (right) runs. Curves are averaged over the final $5000\,r_g/c$. The agreement is excellent for $r\gtrsim5\,r_g$, with modest deviations at smaller radii where strong-field effects become important.}}
\label{fig:appendix_gravitational_torque}
\end{figure*}

Although the weak-field check supports our classification, there remains a possibility that some terms in our formulation are misclassified in the high-spin limit. We therefore compute $\Omega_{\rm LT}$ to higher order using Kerr orbital frequencies \citep{Okazaki1987,Merloni1999}
\begin{equation}
    \Omega_{\phi} = \pm \frac{1}{r^{3/2} \pm a}; \quad
\Omega_{\theta} = \pm \left(\Omega_{\phi}^2 \left(1 \mp \frac{4 a}{r^{3/2}} + \frac{3 a^2}{r^2}\right)\right)^{1/2}; \quad 
\Omega_{LT} \equiv \Omega_{\phi} - \Omega_{\theta},
\end{equation}
where the upper/lower sign refers to prograde/retrograde spin. Figure \ref{fig:appendix_comparing_tilt_of_disk_with_total} compares the resulting LT torque to our derived gravitational torque (averaged over the final $5000\,r_g/c$) for high-spin prograde and retrograde runs. Agreement remains excellent for $r\gtrsim5\,r_g$, with small near-horizon deviations attributable to strong curvature. 

\subsection{Torques in disk frame}
\label{sec:Appendix_Torque equations_in_DF}
Because the disk is tilted, alignment and precession are controlled by the components of the torque projected parallel and perpendicular to the \emph{instantaneous} angular-momentum direction. To avoid geometric mixing of components in the simulation frame, we evaluate the local dynamics of each annulus in a disk-aligned coordinate system $(r,\theta',\phi')$ obtained by a rotation that aligns $\hat{\boldsymbol z}'$ with the shell-integrated angular-momentum unit vector, 
\begin{equation} \hat{\boldsymbol z}' \equiv \hat{\boldsymbol l}(r,t),\qquad \hat{\boldsymbol l}(r,t) \equiv \frac{\left[\boldsymbol L_{\rm total}\right]_{S_r}}{\left|\left[\boldsymbol L_{\rm total}\right]_{S_r}\right|}. 
\end{equation} 
We define the total twist (precession) angle as $\mathcal{P}(r,t)\equiv \tan^{-1}\!\left(\frac{[L_y]_{S_r}}{[L_x]_{S_r}}\right)$, and complete the orthonormal triad by choosing $\hat{\boldsymbol x}'$ to lie in the local averaged-disk plane and sets the zero of $\phi'$,
\begin{equation} 
\hat{\boldsymbol x}' \equiv \frac{\partial \hat{\boldsymbol l}}{\partial \mathcal{I}_{+z}}, \qquad \hat{\boldsymbol y}' \equiv \hat{\boldsymbol z}'\times \hat{\boldsymbol x}' = \frac{1}{\sin\mathcal{I}_{+z}}\frac{\partial \hat{\boldsymbol l}}{\partial \mathcal{P}} , 
\end{equation} 
where $\mathcal{I}_{+z}(r,t)$ is the \emph{instantaneous} tilt relative to the $+z$ axis of the simulation frame. The corresponding angular coordinates $(\theta',\phi')$ are smooth functions of $(\theta,\phi)$ and $(\mathcal{I}_{+z},\mathcal{P})$; explicit expressions are given by \citet{White_2019}. On hypersurfaces of fixed $r$ and $t$, this transformation is a local rotation, so $dr'=dr$ and the radial area element is invariant: \begin{equation} dS'_r \equiv \sqrt{-g'}\,d\theta'\,d\phi' = \sqrt{-g}\,d\theta\,d\phi \equiv dS_r, \end{equation} with $g'\equiv \det(g_{\mu'\nu'})$ and $g_{\mu'\nu'}=\frac{\partial x^\alpha}{\partial x^{\mu'}}\frac{\partial x^\beta}{\partial x^{\nu'}}\,g_{\alpha\beta}$. 

Since $\nabla_\mu T^{\mu}{}_{\nu}=0$ is covariant, the energy--momentum conservation equations and the resulting angular-momentum balance relations have the same form in the disk frame as in the simulation frame after the replacement $(\theta,\phi)\rightarrow(\theta',\phi')$. For reproducibility, we write the full expressions below in a compact shorthand notation as: 
\begin{equation}
\label{eq:appendix_Lzprimetorquefinal}
     \partial_r\left(G_{z^\prime}\right) = \underbrace{-\partial_r \left( \int T^{r}_{\phi^\prime}~dS^\prime_r\right)}_{\text{Matter + EM torque ($z^\prime$-comp)}},
\end{equation}
\begin{equation}
\label{eq:appendix_Lxprimetorquefinal}
\begin{split}
     \partial_r\left(G_{x^\prime}\right) &= \underbrace{\partial_r \left( \int\left(\sin{\phi^\prime}~T^r_{\theta^\prime} + \cot{\theta^\prime} \cos{\phi^\prime}~T^r_{\phi^\prime}\right)~dS^\prime_r\right)}_{\text{Matter + EM torque ($x^\prime$-comp)}} ~+ \\ 
    &\underbrace{\int \left(-\sin{\phi^\prime}~T^{\kappa^\prime}_{\lambda^\prime} \Gamma^{\lambda^\prime}_{\theta^\prime \kappa^\prime} + \cot{\theta^\prime}\sin{\phi^\prime}~T^{\phi^\prime}_{\phi^\prime} + \frac{\cos{\phi^\prime}}{\sin^2{\theta^\prime}} T^{\theta^\prime}_{\phi^\prime} - \cos{\phi^\prime}~T^{\phi^\prime}_{\theta^\prime}\right) dS^\prime_r}_{\text{Gravitational torque ($x^\prime$-comp)}},
\end{split}
\end{equation}
\begin{equation}
\label{eq:appendix_Lyprimetorquefinal}
\begin{split}
     \partial_r\left(G_{y^\prime}\right) &= \underbrace{\partial_r \left( \int \left(-\cos{\phi^\prime}~T^r_{\theta^\prime} + \cot{\theta^\prime} \sin{\phi^\prime}~T^r_{\phi^\prime}\right)~dS^\prime_r\right)}_{\text{Matter + EM torque ($y^\prime$-comp)}} ~+ \\ 
    &\underbrace{\int \left(\cos{\phi^\prime}~T^{\kappa^\prime}_{\lambda^\prime} \Gamma^{\lambda^\prime}_{\theta^\prime \kappa^\prime} - \cot{\theta^\prime}\cos{\phi^\prime}~T^{\phi^\prime}_{\phi^\prime} + \frac{\sin{\phi^\prime}}{\sin^2{\theta^\prime}} T^{\theta^\prime}_{\phi^\prime} - \sin{\phi^\prime}~T^{\phi^\prime}_{\theta^\prime}\right) dS^\prime_r}_{\text{Gravitational torque ($y^\prime$-comp)}},
\end{split}
\end{equation}
where $T^{\mu\prime}_{~\nu^\prime}, \Gamma^{\lambda^\prime}_{\theta^\prime \kappa^\prime}$ are the stress-energy and connection coefficient components in the disk-aligned frame. It is important to note that the Christoffel connection does not transform as a tensor. Physically, this transformation ``unwarps'' each spherical shell by rotating into the local disk orientation. In this frame the torque components have a direct kinematic interpretation: $G_{x'}$ and $G_{y'}$ control changes in the annulus orientation (tilt and twist) with a subtle difference that we describe later, while $G_{z'}$ changes the shell-integrated angular-momentum magnitude and describes accretion torque \citep{Balbus1998}.

Because we diagnose tilt and twist for a \emph{single annulus}, we work with shell–integrated angular–momentum components whose evolution is set by radial fluxes determined by radial torque gradients (\cite{Sorathia_2014}, equations \ref{eq:appendix_Lztorquefinal},\ref{eq:appendix_Lxtorquefinal} and \ref{eq:appendix_Lytorquefinal}). This ``per-shell'' viewpoint introduces an additional, purely geometric contribution when determining the rate of change of tilt and twist: the local disk basis varies with radius, so projections change both because the torque changes and because basis rotate across the shell. Using equation~\ref{eq:totalinclinationeq} (and $\mathcal{I}_{+z}=\mathcal{I}_{\rm BH}$ for prograde, $\mathcal{I}_{+z}=\pi-\mathcal{I}_{\rm BH}$ for retrograde), the simulation-frame expression is
\begin{equation}
\label{eq:appendix_didt_inBHframetorques}
    \partial_t(\mathcal{I}_{+z}) = \frac{1}{\left|\left[\mathbf{L}\right]_{S_r}\right|} \left(\cos{\mathcal{I}_{+z}}\cos{\mathcal{P}} \partial_r\left(G_x^{\text{total}}\right) + \cos{\mathcal{I}_{+z}}\sin{\mathcal{P}} \partial_r\left(G_y^{\text{total}}\right) - \sin{\mathcal{I}_{+z}} \partial_r\left(G_z^{\text{total}}\right)\right)
\end{equation}
In the disk frame the evolution separates cleanly into torque-gradient and geometric (shear-coupling) terms:
\begin{equation}
\label{eq:appendix_didt_indiskframetorques}
    \partial_t(\mathcal{I}_{+z}) = \frac{1}{\left|\left[\mathbf{L}\right]_{S_r}\right|} \left(\partial_r\left(G_{x^\prime}^{\text{total}}\right) - \mathcal{S}_{x^\prime}\right)
\end{equation}
\begin{equation}
\label{eq:appendix_dtwistdt_indiskframetorques}
    \partial_t(\mathcal{P}) = \frac{1}{\left|\left[\mathbf{L}\right]_{S_r}\right| \sin{\mathcal{I}_{+z}}} \left(\partial_r\left(G_{y^\prime}^{\text{total}}\right) - \mathcal{S}_{y^\prime}\right)
\end{equation}
where $\mathcal{S}_{x^\prime}$, $\mathcal{S}_{y^\prime}$ are shear-coupling terms arise from radial variation of the local basis vectors. These equations are similar to \cite{Philippov2014} with the only difference of additional shear-coupling terms. Projecting the simulation–frame (matter+EM) torque components into the disk frame and differentiating the basis vector gives:
\begin{equation}
\label{eq:appendix_sheartermxprime}
    \mathcal{S}_{x^\prime} \equiv \left[\boldsymbol{G}^{\prime, \text{MA+EM}} \cdot \partial_r(\hat{\boldsymbol{x}}^\prime)\right] = - G_{z^\prime} \frac{\partial (\mathcal{I}_{+z})}{\partial r} + G_{y^\prime}^{\text{MA+EM}} \cos{\mathcal{I}_{+z}}\frac{\partial \mathcal{P}}{\partial r} 
\end{equation}
\begin{equation}
\label{eq:appendix_sheartermyprime}
    \mathcal{S}_{y'} \equiv \left[\boldsymbol{G}^{\prime,{\rm MA+EM}}\cdot \partial_r(\hat{\boldsymbol{y}}^\prime)\right]
= -\,\frac{\partial\mathcal{P}}{\partial r}\left(\sin\mathcal{I}_{+z}\,G_{z'} + \cos\mathcal{I}_{+z}\,G_{x'}^{\rm MA+EM}\right)
\end{equation}
with ``MA+EM'' denoting the matter and EM torques from equations \ref{eq:appendix_Lxprimetorquefinal} and \ref{eq:appendix_Lyprimetorquefinal}. The above shear terms (i) vanish for a locally rigid annulus ($\partial_r(\mathcal{I}_{+z}) = \partial_r(\mathcal{P}) = 0$), (ii) are independent of coordinate gauge because they are inner products; and (iii) do not represent a new physical torque channel. Rather, they are connection-like terms that encode the coupling between the physical torque and the radially rotating, non-inertial disk frame (analogous to Coriolis-type terms in rotating coordinates).

Finally, to leading order in $a/r$ we estimate the gravitational torque in the disk frame by projecting the LT angular velocity onto the local basis,
\begin{equation}
\label{eq:appendix_disk_LTvector}
\boldsymbol{\Omega}'_{LT}(r,t) \;=\; \big(-\,\Omega_{LT}\,\sin\mathcal{I},\; 0,\; \Omega_{LT}\,\cos\mathcal{I}\big)\, .
\end{equation}
so that the per-shell torque satisfies
\begin{equation}
\label{eq:appendix_tauprimeG_def}
\partial_r(\boldsymbol{G}^{^\prime}_{\text{gravity}}) \approx \boldsymbol{\Omega}'_{LT}\times (|\boldsymbol L|_{S_r}\,\hat{\boldsymbol z}^\prime)
\end{equation}
implying
\begin{equation}
\label{eq:appendix_tauprimeG_components}
\partial_r\!\left(G^{\text{gravity}}_{x'}\right) \;\simeq\; 0,\qquad
\partial_r\!\left(G^{\text{gravity}}_{y'}\right) \;\simeq\; \Omega_{LT}\,|\boldsymbol L|_{S_r}\,\sin\mathcal{I}_{+z},\qquad
\partial_r\!\left(G^{\text{gravity}}_{z'}\right) \;\simeq\; 0 \quad (r\gg a)\, .
\end{equation}
As expected, the LT torque is perpendicular to $\boldsymbol L$ and therefore drives precession without changing $|\boldsymbol L|_{S_r}$ or $\mathcal{I}_{+z}$. Since $\sin\mathcal{I}_{+z}>0$ and $\Omega_{\rm LT}$ has the sign of the BH spin (positive for prograde, negative for retrograde), LT precession is always in the sense set by the spin

\section{Measurement of the jet inclination}
\label{sec:appendix_jet_inclination}
We measured jet inclination using the method of \cite{Liska2018jetprecess}, with one modification: we identify the jet by magnetization, $\sigma \equiv b^2/(2\rho) > 1$, rather than their criterion $p_{\rm mag}/\rho > 1/(2r)$. The two prescriptions agree within $\sim 50\,r_g$, but at larger radii the \cite{Liska2018jetprecess} criterion aligned more closely with the disk due to increased mass loading. For $r>10\,r_g$ we separate the jet into upper ($0<\theta<\pi/2$) and lower ($\pi/2<\theta<\pi$) hemispheres. For $r\le 10\,r_g$, where the tilted disk complicates a hemispheric split, we instead use the sign of the radial magnetic field: upper ($b^r>0$) and lower ($b^r<0$). At each $(r,t)$ we compute magnetic-pressure-weighted centroids of the upper and lower regions in Cartesian coordinates and define the local jet axis from the centroid separation; the reported jet inclination is obtained by averaging this axis over radius.

\section{Symmetry and Tilt Consistency in Prograde and Retrograde MADs}
\label{sec:Appendix_Symmetry and Tilt Consistency in Prograde and Retrograde MADs}
Because only the \emph{relative} orientation between the BH spin and disk angular momentum is physical, a prograde/retrograde setup can be implemented either by rotating the torus at fixed spin or by reversing the spin at fixed torus orientation. To verify that this symmetry does not bias the MAD state or the tilt evolution, we performed two additional runs. For the physically prograde configuration we set $a=-0.9375$ and rotate the torus angular momentum by $164^\circ$ about the $y$-axis, giving an effective misalignment of $16^\circ$ between the BH spin and disk. For the physically retrograde configuration we set $a=+0.9375$ with the same $164^\circ$ torus rotation, giving a misalignment of $164^\circ$. Both runs use resolution $160\times128\times80$ in $(r,\theta,\phi)$, and disk tilt is computed using equation~\ref{eq:diskinclinationeq}.
\begin{figure*}[!ht]
\centering
\includegraphics[width=\textwidth]{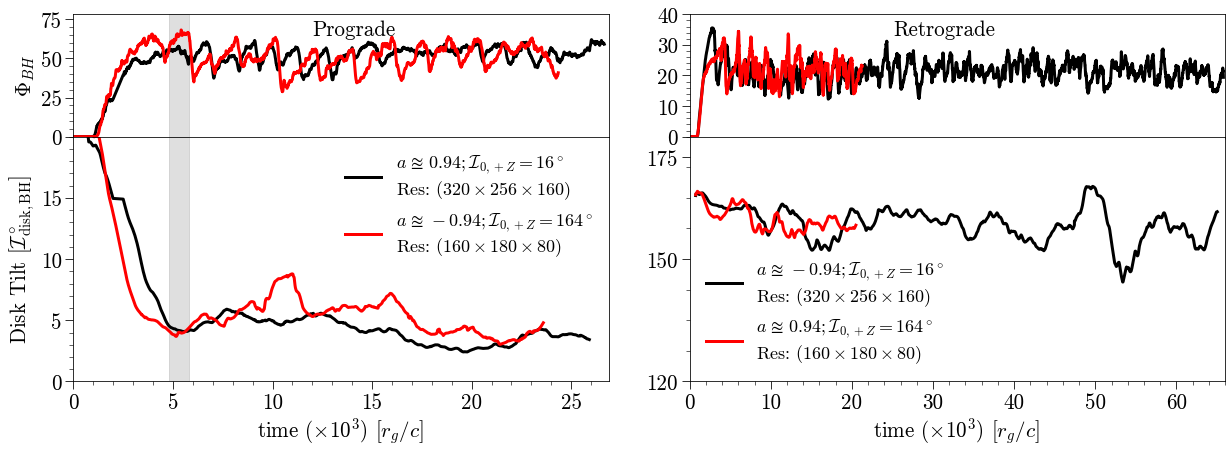}
\caption{\small{Test of the setup symmetry between BH-spin sign and initial torus rotation. Curves compare the horizon magnetic-flux and the inner-disk tilt (measured out to $20\,r_g$ using equation~\ref{eq:diskinclinationeq}) between physically equivalent realizations of prograde (left column) and retrograde (right column) configurations. Black curves show the reference high-spin runs ($a\simeq0.94$, $\mathcal{I}_{0,+z}=16^\circ$ for prograde; $a\simeq-0.94$, $\mathcal{I}_{0,+z}=16^\circ$ for retrograde), while red curves show the alternative implementations obtained by flipping the spin sign and adjusting the torus rotation. The MAD saturation and tilt evolution agree closely, indicating that the results are insensitive to this setup convention (and remain consistent when the resolution is reduced by a factor of eight).}}
\label{fig:appendix_tilt_BHspin_symmetry}
\end{figure*}

Figure~\ref{fig:appendix_tilt_BHspin_symmetry} shows that both the MAD level and the inner ($\le 20\,r_g$) tilt evolution are consistent between symmetry-related setups for the prograde and retrograde cases. Although reversing the spin sign changes the ISCO in \textsc{HARMPI} and can modify near-horizon dynamics, the tilt evolution exhibits similar trends and magnitudes across configurations, indicating that the tilt dynamics are robust to this setup choice.
\bibliography{referencesfile}{}
\bibliographystyle{aasjournal}

\end{document}